\documentclass[aps,prl,twocolumn,amsmath,amssymb,nofootinbib,superscriptaddress]{revtex4-2}
\usepackage{adjustbox}
\usepackage[utf8]{inputenc} 
\usepackage{hyperref}       

\usepackage{url}            
\usepackage{booktabs}       
\usepackage{amsfonts}       
\usepackage{nicefrac}       
\usepackage{microtype}      
\usepackage{bm}
\usepackage{bbm}
\usepackage{braket}
\usepackage{graphicx}
\usepackage{subfig}
\usepackage{float}
\usepackage{amsthm}
\usepackage{algorithmic}
\usepackage[linesnumbered,ruled]{algorithm2e}
\SetKwInOut{Parameter}{parameter}

\usepackage{mathtools}
\usepackage{nccmath}
\usepackage{caption}
\usepackage{multirow}
\usepackage{ulem}
\usepackage{titlesec}

\usepackage{soul}
\titlespacing{\title}{0pc}{0.1pc}{0.3pc}
\titlespacing{\section}{0pc}{0.1pc}{0.3pc}
\newcommand{\expect}[2]{\left\langle #1 \left| #2 \right| #1 \right\rangle}
\newcommand{\avg}[1]{\left\langle #1 \right\rangle}

\usepackage{changes}

\newcommand{\affa}{Beijing Key Laboratory of Fault-Tolerant Quantum Computing, \\Beijing Academy of Quantum Information Sciences, Beijing 100193, China}
\newcommand{\affb}{Beijing National Laboratory for Condensed Matter Physics, \\Institute of Physics, Chinese Academy of Sciences, Beijing 100190, China}
\newcommand{\affc}{School of Physical Sciences, University of Chinese Academy of Sciences, Beijing 100190, China}
\newcommand{\affd}{Hefei National Laboratory, Hefei 230088, China}

\begin{document}

\title{Observing Quantum Correlation Dynamics in Tunable Superconducting Bose-Hubbard Simulators}

\author{Z. T. Wang}
\thanks{These authors contributed equally to the work.}
\address{\affa}
\author{Si-Yun Zhou}
\thanks{These authors contributed equally to the work.}
\address{\affb}
\address{\affc}
\author{Yun-Hao Shi}
\thanks{These authors contributed equally to the work.}
\address{\affb}
\author{Kaixuan Huang}
\thanks{These authors contributed equally to the work.}
\address{\affa}
\author{Z. H. Yang}
\address{\affb}
\address{\affc}
\author{Jingning Zhang}
\email{zhangjn@baqis.ac.cn}
\address{\affa}
\author{\mbox{Kui Zhao}}
\address{\affa}
\author{Yueshan Xu}
\address{\affa}
\author{Hao Li}
\address{\affa}
\author{S. K. Zhao}
\address{\affa}
\author{Yulong Feng}
\address{\affa}
\author{Guangming Xue}
\address{\affa}
\author{Yu Liu}
\address{\affb}
\address{\affc}
\author{Wei-Guo Ma}
\address{\affb}
\address{\affc}
\author{\mbox{Cai-Ping Fang}}
\address{\affb}
\address{\affc}
\author{Hao-Tian Liu}
\address{\affb}
\address{\affc}
\author{Yong-Yi Wang}
\address{\affb}
\address{\affc}
\author{Kai Xu}
\address{\affb}
\address{\affa}
\address{\affd}
\author{Haifeng Yu}
\email{hfyu@baqis.ac.cn}
\address{\affa}
\address{\affd}
\author{Heng Fan}
\email{hfan@iphy.ac.cn}
\address{\affb}
\address{\affa}
\address{\affd}
\author{S. P. Zhao}
\email{spzhao@iphy.ac.cn}
\address{\affa}
\address{\affb}

\date{\today}

\pacs{xxx}

\begin{abstract}

The dynamics of quantum correlations are central to understanding many physical properties of quantum systems. Here we experimentally study the correlation dynamics via two-particle quantum walks in superconducting Bose-Hubbard qutrit arrays, with tunable on-site interaction $U$ realized by Floquet engineering. Quantum walks show the characteristic change from bosonic bunching to fermionic antibunching with increasing $U$. The two-site entanglement and quantum correlation dynamics, as measured by negativity and quantum discord, are investigated. We find that depending on the initial state, the propagation of entanglement can be strongly suppressed with increasing $U$, while that of quantum discord exhibits considerably larger amplitude; or both of them appear insensitive to $U$. Furthermore, the forms of entanglement are found to persist throughout particle walks for $U =$ 0 and it is generally not the case when $U$ increases. Our work highlights the role of interaction in shaping quantum dynamics and extends the realm of simulating correlated quantum systems with superconducting circuits.

\end{abstract}

\maketitle

\newcommand{\ketbra}[2]{\left|#1\right\rangle\left\langle#2\right|}

{\it Introduction.}---Quantum walks are a useful tool in the studies of particle dynamics~\cite{lahini_quantum_2012, schreiber_2d_2012, peruzzo_quantum_2010, karski_quantum_2009, fukuhara_microscopic_2013, preiss_strongly_2015, yan_strongly_2019, gong_quantum_2021, giri_signatures_2022}, topology~\cite{weidemann_topological_2022, lin_topological_2022, xiao_observation_2024, zhang_nonchiral_2025}, anyon physics~\cite{kwan_realization_2024}, and simulation of many-body systems~\cite{braumuller_probing_2022}, which span many physical platforms like photons~\cite{lahini_quantum_2012, peruzzo_quantum_2010, schreiber_2d_2012, weidemann_topological_2022, lin_topological_2022, xiao_observation_2024, zhang_nonchiral_2025}, neutral atoms~\cite{karski_quantum_2009, fukuhara_microscopic_2013, preiss_strongly_2015, kwan_realization_2024}, and superconducting qubits~\cite{yan_strongly_2019, gong_quantum_2021, braumuller_probing_2022}. Among these studies, the Hanbury-Brown-Twiss interference and fermionization have been demonstrated in the Bose-Hubbard (BH) model~\cite{lahini_quantum_2012, preiss_strongly_2015, yan_strongly_2019}, and the Bloch oscillations, Mach-Zehnder interferometer, and coherent dynamics have been explored~\cite{lahini_quantum_2012, schreiber_2d_2012, fukuhara_microscopic_2013, preiss_strongly_2015, yan_strongly_2019, gong_quantum_2021}. The two-particle dynamics have been of particular interest, since they may serve as the building blocks and a valuable starting point for the bottom-up studies of many-body quantum dynamics~\cite{lahini_quantum_2012, schreiber_2d_2012, fukuhara_microscopic_2013}. Experimentally, the studies using photons~\cite{lahini_quantum_2012, peruzzo_quantum_2010, schreiber_2d_2012} and neutral atoms~\cite{karski_quantum_2009, fukuhara_microscopic_2013, preiss_strongly_2015} are often based on the measurements of the probability distributions and correlation functions, and the model parameters, such as the on-site interaction $U$ and tunneling strength $J$ in the BH model, can be conveniently tuned~\cite{lahini_quantum_2012, preiss_strongly_2015}. For the superconducting circuits, on the other hand, the multi-qubit~\cite{song_10-qubit_2017} and qudit~\cite{liu_performing_2023, vbh4-lysv} density matrices can be further obtained via particle number non-conserving measurements, thus providing more details in experiment, but $U$ is usually not tunable for the widely used transmon devices in simulating the BH model.

\begin{figure*}[t]
	\centering
	\includegraphics[width=0.95\linewidth]{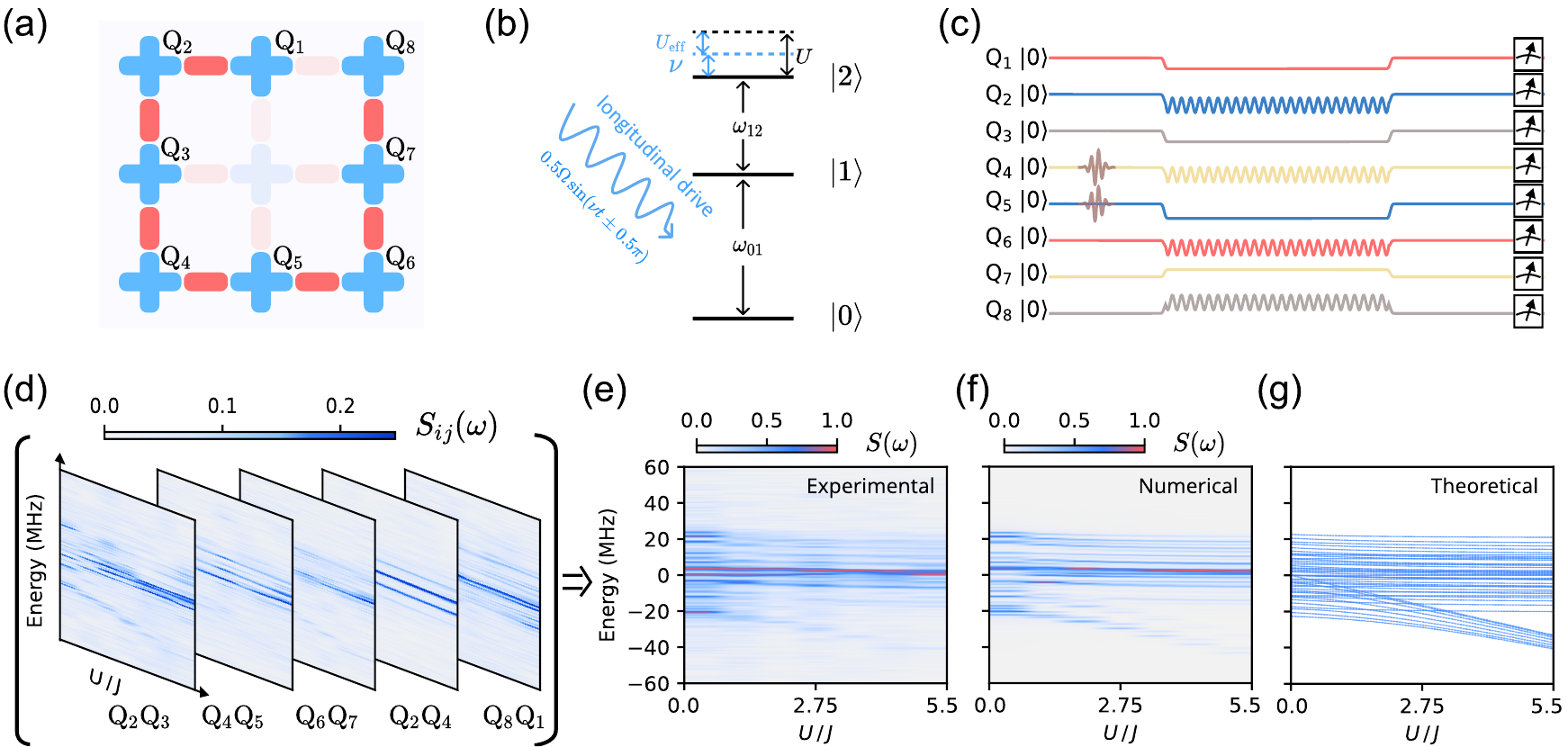}
	\caption{\small{\textbf{Experiment and energy spectrum measurement.} (a) Schematic superconducting processor. Eight qutrits (crosses) are selected to form a chain by setting the couplers (rectangles) in different states. (b) Qutrit level shifts under longitudinal drive. (c) The measurement of partial energy spectrum. Two qutrits Q$_4$ and Q$_5$ are both initialized to the  $|+\rangle$ state. All qutrits are brought to the resonant frequency to evolve with longitudinal drives applied to the even-labeled qutrits. At the end of evolution, each qutrit is measured in the Pauli-X or Pauli-Y basis.  (d) Partial energy spectra reflected by Fourier transform magnitudes $S_{ij}(\omega)$ for five qutrit pairs. (e) Total energy spectrum of the system obtained by summing up the partial spectra in (d). (f) Spectrum by numerical simulation of experiment. (g) Spectrum by exact diagonalization.}}
	\label{fig1}
\end{figure*}

In this work, we present an experimental study of two-particle (-site) correlation dynamics via quantum walks in the superconducting qutrit arrays. We use a novel Floquet engineering approach to realize the BH model with fully tunable $U$ and, at the same time, the measurements of single- and two-qutrit density matrices. These enable the studies of the dynamics of the two-site entanglement and quantum discord~\cite{RevModPhys.84.1655,Modi2012, HU20181} that are widely used to identify quantum states in many-body equilibrium~\cite{osborne_entanglement_2002, osterloh_scaling_2002, dillenschneider_quantum_2008, sarandy_classical_2009, allegra_quantum_2011} and nonequilibrium~\cite{heyl_dynamical_2018, vidal_entanglement_2004, sende_dynamical_2005, mishra_tuning_2013, mishra_survival_2016, campbell_dynamics_2017, mishra_dynamical_2018, vimal_entanglement_2024} phase diagrams. Quantum discord is originally introduced to describe quantum correlations beyond entanglement~\cite{Zurek_2001}, whose study has been limited to two-level systems due to the computational complexity with increasing Hilbert space dimension ~\cite{de_chiara_genuine_2018}. Recently, it is explored in a two qutrit spin system, demonstrating its existence in the absence of entanglement~\cite{fu_experimental_2022}. Here we extend the study to the three-level BH arrays, focusing on the dynamics of the two-site entanglement and quantum discord. We find that their dynamics can behave in very different ways depending on the initial states and $U$. While entanglement propagation is strongly suppressed, quantum discord can propagate with considerably larger amplitude. As identified from the density matrices, the forms of entanglement for $U =$ 0 will not change from one qutrit pair to the next during particle walks. Generally, this feature no longer exists as $U$ increases but can be seen in some cases such as quantum walks with the Bell initial state.

{\it Experimental setup and protocol.}---Our experiments are performed on a superconducting processor with 9 transmon qutrits and 12 tunable couplers arranged in a square lattice~\cite{yang_mitigation_2024, wang_demonstration_2024}, as shown in Fig.~\ref{fig1}(a) (see also Supplemental Material~\cite{supplementary}). After decoupling the couplers by the Schrieffer-Wolff transformation~\cite{PhysRevApplied.10.054062}, an $L$-qutrit array ($L$ is the total qutrit number) is described by the BH model with the parameters of the $l$-th qubit frequency $\omega_l$, anharmonicity $\alpha_l$, and nearest-neighbor coupling $g$~\cite{10.1063/1.5089550, Roushan2017}. Since transmons typically have fixed $\left|\alpha_l\right|\gg g$, the system is in the strong interaction regime close to the hard-core limit, where double occupancies are strongly suppressed. 

We consider an $L$-qutrit chain and use Floquet engineering to modulate the frequencies of even-labeled qutrits according to $\omega_l(t)=\omega_{\rm res}+\Omega\cos\nu t$, as shown in Figs.~\ref{fig1}(b) and (c). Here, $\omega_{\rm res}$ is the resonant frequency common for all qubits during time evolution, and $\Omega$ and $\nu$ are the amplitude and frequency of the longitudinal drive, respectively. Similar technique has been used to tune the coupling strength between qubits~\cite{cai2019observation, zhao_probing_2022, shi2023}, where the frequency of longitudinal field is far detuned from the anharmonicity. In order to tune the on-site interaction, the frequency of the longitudinal fields is in the vicinity of anharmonicity, which effectively shifts the energy of the doubly-excited states. Under the condition ${\mathcal J}_0\left(\frac{\Omega}{\nu}\right)={\mathcal J}_1\left(\frac{\Omega}{\nu}\right)$ with ${\mathcal J}_m$ being the $m$-th order Bessel function, so that $\Omega/\nu \approx 1.4347$ is required, the dynamics is governed by the BH model, valid within the two-particle subspace:
\begin{eqnarray}
\hat H_{\rm BH}=J\sum_{l=1}^{L-1}\left(\hat a_l^\dagger\hat a_{l+1}+h.c.\right)-\frac{U}{2}\sum_{l=1}^{L}\hat a_l^\dagger\hat a_l^\dagger\hat a_l\hat a_l~,~~~ \label{eq:H_BH}
\end{eqnarray}
where $U=\left|\nu+\alpha_l\right|$, $J=g{\mathcal J}_0\left(\frac{\Omega}{\nu}\right)$, and $\hat a_l^\dagger$ ($\hat a_l$) is the bosonic creation (annihilation) operator~\cite{supplementary}. We note that $\alpha_l$ characterized by two-photon excitation measurements is slightly different from qutrit to qutrit while $\nu$ is uniform, so $U$ also slightly varies, which is not reflected in Eq.~(\ref{eq:H_BH}) for simplicity~\cite{remark}.

\begin{figure*}[t]
	\centering
	\includegraphics[width=0.99\linewidth]{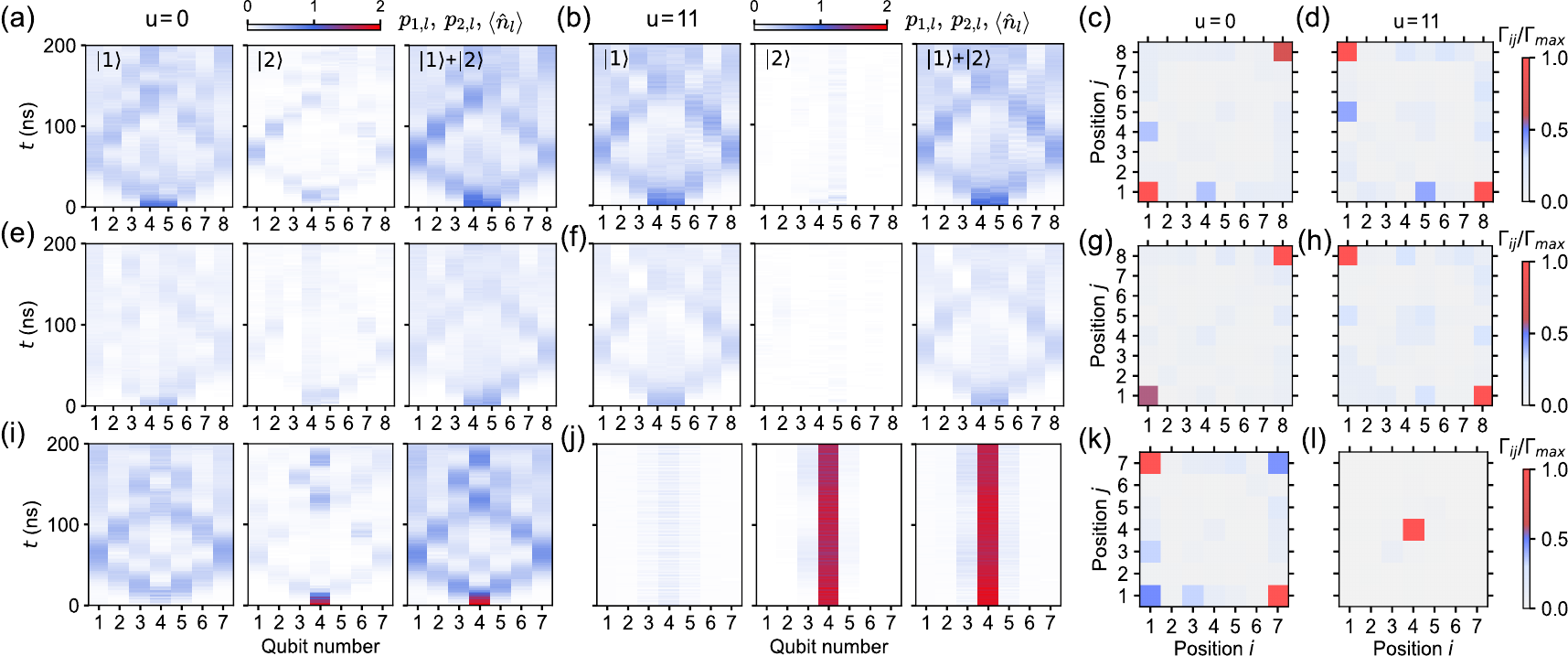}
	\caption{\small{\textbf{Quantum walks and density-density correlations with different initial states and $u$.} (a)–(d) Experimental results for the initial state $\ket{\Phi_1}$, i.e., an 8-qutrit chain (Q$_1$ to Q$_8$) with two particles initially placed in the middle two qutrits Q$_4$ and Q$_5$. (e)–(h) Corresponding results with the initial state $\ket{\Phi_2}$. (i)-(l) Experimental results for the initial state $\ket{\Phi_3}$, i.e., a 7-qutrit chain (Q$_1$ to Q$_7$) with two particles initially placed in the middle qutrit Q$_4$. (a), (b), (e), (f), (i), (j) Quantum walks. (c), (d), (g), (h), (k), (l) Density-density correlations measured at $t$ = 68 ns for (c), (d), (g), (h) and 64 ns for (k), (l).}}
	\label{fig2}
\end{figure*}

{\it Energy spectrum measurement.}---We experimentally measure the energy spectrum of the system~\cite{shi2023,Xiang2023,Roushan2017} to show the tunability of $u=U/J$, using an 8-qutrit chain realized with different couplers in coupling or decoupling states as shown by the dark and light orange rectangles in Fig.~\ref{fig1}(a), respectively. The qutrits have $\alpha_l$ ranging from $-201$ to $-211$ MHz and the coupling strength is set to be $g=11$ MHz. By applying the longitudinal drive, we have a fixed $J=g{\mathcal J}_0\left(1.4347\right)\simeq 6.02$ MHz, while $u$ can be tuned with $\nu$~\cite{supplementary}. The experiment starts from an initial state $\ket{\phi_{ij}}$ with a pair of qutrits ${\rm Q}_i$ and ${\rm Q}_j$ each prepared in $\ket{+}_l\equiv\left(\ket{0}_l+\ket{1}_l\right)/\sqrt{2}$ ($l\in\left\{i,j\right\}$). The system then evolves under longitudinal drives with $\omega_{\rm res}$ = 4.5 GHz, as shown in Fig.~\ref{fig1}(c) for $i$=4, $j$=5 as an example. We measure the Pauli observables $\hat\sigma_i^\alpha\hat\sigma_j^\beta$ ($\alpha\in\{x,y\}$) at 160 points over 800 ns as time evolves. The Fourier power spectrum $S_{ij}(\omega)=|{\mathcal F}_\omega\left[\chi_{ij}(t)\right]|$ of the time-dependent correlation function $\chi_{ij}(t)=\expect{\psi_{ij}(t)}{\hat\sigma_i^+\hat\sigma_j^+}$, with $\hat\sigma_l^+=\left(\hat\sigma_l^x+\mathrm{i}\hat\sigma_l^y\right)/2$ and $\ket{\psi_{ij}(t)}=\exp\left(-\mathrm{i}\hat H_{\rm BH}t\right)\ket{\phi_{ij}}$, is shown in Fig.~\ref{fig1}(d) for five qutrit pairs. Each of them contains partial information about the two-particle spectrum~\cite{supplementary}. The total energy spectrum is summed up as $S(\omega)=\sum_{i,j}S_{ij}(\omega)$ shown in Fig.~\ref{fig1}(e), where the eigenenergies are located by the high-density ridges. Figures~\ref{fig1}(f) and (g) show the results obtained by numerical simulation of the experiment and exact diagonalization calculation of the Hamiltonian Eq.~(\ref{eq:H_BH}). Although it is hard to discriminate each of the eigenenergies in Figs.~\ref{fig1}(e) and (f) due to the limited evolution time, it can be clearly observed that a two-band structure emerges from a single continuous band as $u$ increases. 

\begin{figure*}[t]
	\centering
	\includegraphics[width=0.99\linewidth]{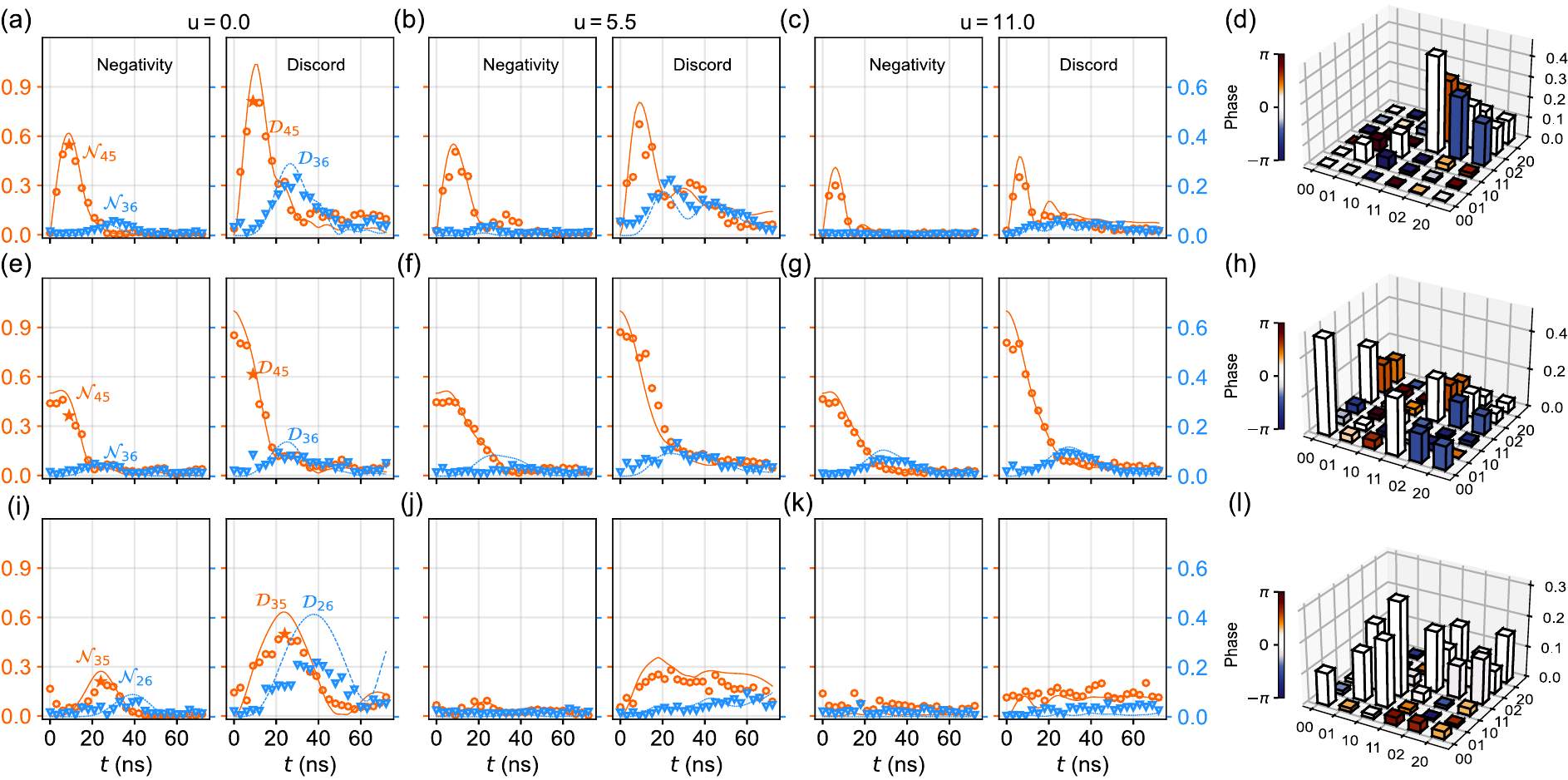}
	\caption{\small{\textbf{Negativity ${\mathcal N}_{ij}$ and quantum discord ${\mathcal D}_{ij}$ with different initial states and $u$.} (a)-(c) ${\mathcal N}_{45}$, ${\mathcal N}_{36}$, ${\mathcal D}_{45}$, and ${\mathcal D}_{36}$ with the initial state $\ket{\Phi_1}$ for the 8-qutrit chain. (d) Density matrix for Q$_4$Q$_5$ at the points shown as stars in (a). (e)-(h) Corresponding results with the initial state $\ket{\Phi_2}$. (i)-(k) ${\mathcal N}_{35}$, ${\mathcal N}_{26}$, ${\mathcal D}_{35}$, and ${\mathcal D}_{26}$ with the initial state $\ket{\Phi_3}$ for the 7-qutrit chain. (l) Density matrix for Q$_3$Q$_5$ at the points shown as stars in (i). Symbols and lines are the experimental results and theoretical results calculated using the experimental parameters, circles (solid lines) and triangles (dashed lines) correspond to the left and right scales, respectively.}}
	\label{fig3}
\end{figure*}

{\it Two-particle quantum walks.}---To study two-particle quantum walks and correlation dynamics under tunable $u$, we consider three different initial states, including $\ket{\Phi_1}=\hat a_4^\dag\hat a_5^\dag\otimes_{i=1}^{8}\ket{0}_i$ and $\ket{\Phi_2}=\ket{B}_{45}\otimes_{i\neq4,5}\ket{0}_i$ with the Bell state $\ket{B}_{45}$ = $\left(\ket{0}_4\ket{0}_5+\ket{1}_4\ket{1}_5\right)/\sqrt{2}$ for an 8-qutrit chain, and $\ket{\Phi_3}=(1/\sqrt{2})\hat a_4^\dag\hat a_4^\dag\otimes_{i=1}^{7}\ket{0}_i$ for a 7-qutrit chain. The walks reflect the time evolution of the average particle number $\avg{\hat n_l}=\sum_nnP_{n,l} \equiv \sum_n p_{n,l}$ ($\hat n_l=\hat a_l^\dag\hat a_l$) of the individual qutrits. Here $P_{n,l}\equiv{\rm tr}\left[\hat\rho_l(t)\ket{n}\bra{n}_l\right]$ and $\rho_l(t)$ is the $l$-th qutrit density matrix. The photon density distribution $p_{n,l}$ can be experimentally read out simultaneously at a given time $t$ for all levels $n$ and all qutrits $l$, as shown in Figs.~\ref{fig2}(a)-(b), (e)-(f), and (i)-(j) for $\ket{\Phi_1}$, $\ket{\Phi_2}$, and $\ket{\Phi_3}$, respectively. We observe that the main effect of increasing $u$ is the suppression of the probability transfer from single- to double-excitation states for $\ket{\Phi_1}$ and $\ket{\Phi_2}$, and is opposite for $\ket{\Phi_3}$. Figures~\ref{fig2}(c) and (d) show the results of density-density correlator $\Gamma_{ij}(t) = \langle \hat a_i^{\dagger}\hat a_j^{\dagger}\hat a_j\hat a_i\rangle_t$ measured at 68 ns when the two particles initially placed at the central two qutrits walk to the edges of the chain. As $u$ increases, the maxima of the correlator change from diagonal to anti-diagonal, indicating a change from bosonic bunching to fermionic antibunching due to the Hanbury-Brown-Twiss interference and quantum statistics~\cite{lahini_quantum_2012, preiss_strongly_2015, yan_strongly_2019}. Figures~\ref{fig2}(e)-(h) show the similar results with lowered particle density when the Bell state $\ket{\Phi_2}$ is prepared at the central two qutrits. This is expected because the Bell state $\ket{\Phi_2}$ contains a component of the state $\ket{\Phi_1}$, while its vacuum component does not contribute to $\Gamma_{ij}(t)$. Figures~\ref{fig2}(k) and (l) show the correlator measured at 64 ns when the two particles are initially placed at one central qutrit. From Figs.~\ref{fig2}(i)-(l), one can see that for $u=0$, the two particles are identical and walk freely in both left and right directions. However, for large $u$ they almost localize as a bound pair restricted by the energy conservation. The propagation of the pair is much slower, described by a much reduced tunneling rate of $2J^2/U \ll J$.

{\it Entanglement and quantum correlation.}---We investigate the dynamics of two-site entanglement and quantum discord~\cite{de_chiara_genuine_2018, Zurek_2001} during quantum walks. Experimentally, we measure the two-qutrit density matrix via quantum state tomography. The entanglement measured by negativity is then calculated from ${\mathcal N}_{ij}=\left(\left\|\hat\rho_{ij}^{\rm PT}(t)\right\|_1-1\right)/2$~\cite{RevModPhys.84.1655}, where $\hat\rho_{ij}^{\rm PT}(t)$ is the reduced density matrix for $i$-th and $j$-th qutrits at time $t$ partially transposed with respect to the $i$-th qutrit, and $\left\|\cdot\right\|$ is the trace norm. The quantum discord is calculated by subtracting classical correlation $C\left(\hat\rho_{ij}\right)$ from quantum mutual information $I\left(\hat\rho_{ij}\right)$, i.e., ${\mathcal D}_{ij}=I\left(\hat\rho_{ij}\right)-C\left(\hat\rho_{ij}\right)$~\cite{Modi2012,HU20181} (see Supplemental Material for details~\cite{supplementary}).

The experimental results for the first two qutrit pairs Q$_4$Q$_5$ and Q$_3$Q$_6$ with the initial state $\ket{\Phi_1}$ are shown in Figs.~\ref{fig3}(a), (b), and (c) for three different $u$ values (symbols), together with the theoretical results calculated using the experimental parameters (lines). In this case, two particles are initially placed in the middle adjacent qutrits Q$_4$ and Q$_5$ in the 8-qutrit chain. We see that ${\mathcal N}_{45}$ develops from zero with increasing $t$ to maximum values and then decreases, indicating the increase and decrease of the entanglement. As $u$ increases, the entanglement seen in ${\mathcal N}_{45}$ weakens. The entanglement propagation to the next Q$_3$Q$_6$ qutrit pair can be clearly seen from ${\mathcal N}_{36}$, which quickly decreases and almost vanishes for larger $u$. On the other hand, the quantum discord ${\mathcal D}_{45}$ exhibits a similar trend as  ${\mathcal N}_{45}$, but its propagation to the next Q$_3$Q$_6$ qutrit pair appears more prominent than that of negativity. These results indicate that a large portion of the quantum correlation is not in the form of entanglement during time evolution, consistent with the observation that nonclassical correlations can exist in state where entanglement is zero~\cite{fu_experimental_2022,Yao2015}.

For a similarly prepared 2-qutrit system $\hat a_1^\dag\hat a_2^\dag\otimes_{i=1}^{2}\ket{0}_i$ with $u=0$, the initial product state $\ket{11}$ is known to evolve periodically to maximally entangled state $\ket{\Psi_2}$ = $\left(\ket{20}+\ket{02}\pm\ket{11}\right)/\sqrt{3}$ and then to the symmetric state $\ket{\Psi^{\prime}_2}$ = $\left(\ket{20}+\ket{02}\right)/\sqrt{2}$ before returning to the initial state and starting re-entangling. Fig.~\ref{fig3}(d) illustrates the density matrix of the Q$_4$Q$_5$ qutrit pair measured at the time indicated by the stars in Fig.~\ref{fig3}(a) for $u$ = 0. The matrix elements are seen to locate dominantly in the two-particle subspace spanned by $\{\ket{11}$, $\ket{02}$, $\ket{20}\}$, entangled in the form similar to $\ket{\Psi_2}$. At later times, they will evolve through the form similar to $\ket{\Psi^{\prime}_2}$. Meanwhile, as the particles leave the Q$_4$Q$_5$ qutrit pair during quantum walk, the population in the single-particle subspace $\{\ket{01}, \ket{10}\}$ may grow incoherently, which makes no contribution to entanglement. For larger $u$,  entanglement in the two-particle subspace will be increasingly suppressed~\cite{supplementary}.

Figures \ref{fig3}(e)-(h) show the results for the initial state $\ket{\Phi_2}$ with the Bell state $\ket{B}_{45}$ initially prepared in the Q$_4$Q$_5$ qutrit pair. The propagation of entanglement and quantum discord seen from ${\mathcal N}_{36}$ and ${\mathcal D}_{36}$ changes moderately in this case for increasing $u$, while the magnitude of entanglement slightly increases, in contrast to the vanishing entanglement in Fig.~\ref{fig3}(c). This results from the Bell-state entanglement, which is absent in the $\ket{\Phi_1}$ state and does not involve the double-excitation state so is independent of $u$. The measured density matrix in Fig.~\ref{fig3}(h) shows that the initial population of the $\ket{11}$ component in the Bell state coherently transfers among the basis states in the two-particle subspace $\{\ket{11},\ket{02},\ket{20}\}$ resulting in entanglement in the form similar to $\ket{\Psi_2}$ (and later to $\ket{\Psi^{\prime}_2}$), while at the same time maintaining the coherence with the $\ket{00}$ component.

The experimental results for the initial state $\ket{\Phi_3}$,  namely with two particles placed in the middle Q$_4$ qutrit in a 7-qutrit chain, are shown in Figs.~\ref{fig3}(i)-(l). The entanglement and quantum discord data for two qutrit pairs Q$_3$Q$_5$ and Q$_2$Q$_6$ show their fast decreases in the measured time scale as $u$ increases, due to the much reduced tunneling rate mentioned above. For $u$ = 0, the entanglement between particles walking in the left and right directions develops in both single- and two-particle subspaces in the forms of $\ket{\Psi_1}$ = $\left(\ket{10}\pm\ket{01}\right)/\sqrt{2}$ and $\ket{\Psi_2}$, as evident from the measured density matrix in Fig.~\ref{fig3}(l). For large $u$, the two particles move together with equal probabilities to the left and right, creating the symmetric entangled state $\ket{\Psi^{\prime}_2}$ over a much longer time scale.   

We have tracked the dynamics of entanglement and quantum discord in the entire qutrit chains including their reflections at the boundaries for the three initial states $\ket{\Phi_1}$, $\ket{\Phi_2}$, and $\ket{\Phi_3}$ and different $u$ values~\cite{supplementary}. It is interesting to note that for $u =$ 0, the forms of entanglement shown in Figs.~\ref{fig3}(d), (h), and (l), namely entangled in the two-particle subspace, similar entanglement plus coherence with the $\ket{00}$ state, and entangled in both single- and two-particle subspaces, would persist throughout the particle walking process. This feature gradually disappears with increasing $u$ as a result of the suppression of population transfer between single- and double-excitation states depending on the initial state. However, in the case of the initial state $\ket{\Phi_2}$, clear signature of the Bell-state entanglement can be seen throughout particle walks for all $u$ values. 

{\it Summary.}---We have performed quantum walk experiments in the BH system with fully tunable $u$ using superconducting qutrit arrays incorporating the Floquet engineering technique. The two-particle (-site) correlation dynamics, including density-density correlation, entanglement, and quantum discord, have been experimentally investigated via particle number non-conserving measurement that is not readily accessible in other platforms. We find that the entanglement and quantum discord can exhibit distinct behaviors during propagation depending on different initial states and changing $u$. In particular, quantum discord, which captures broader nonclassical correlations, can propagate in the three-level BH systems when entanglement almost vanishes. Throughout the quantum walk process, the forms of the entanglement can be preserved during propagation from one qutrit pair to the next. These results provide valuable insights for the future studies of correlated quantum systems in generic many-body scenarios~\cite{lahini_quantum_2012, schreiber_2d_2012, fukuhara_microscopic_2013, preiss_strongly_2015}.

{\it Acknowledgments.}---We thank M.-L. Hu and T. Ma for the helpful discussions. This work was supported by the National Natural Science Foundation of China (Grant Nos. 12504593, T2121001, 92265207, 92365206, T2322030, 92365301, 12504576 and 12404578), the Innovation Program for Quantum Science and Technology (Grant No. 2021ZD0301800), the Beijing Nova Program (Grant No. 20220484121), the Beijing National Laboratory for Condensed Matter Physics (Grant No. 2024BNLCMPKF022), Young Elite Scientists Sponsorship Program of the Beijing High Innovation Plan (Grant No. 20250945),
Beijing Natural Science Foundation (Grant No. 1262048), and the China Postdoctoral Science Foundation (Grant No. GZB20240815).

{\it Data availability.}---The data that support the findings of this article are openly available ~\cite{datasetdoi}.

%

\end{document}


\title{Supplemental Material for ``Observing Quantum Correlation Dynamics in Tunable Superconducting Bose-Hubbard Simulators"}
	
\author{Z. T. Wang}
\thanks{These authors contributed equally to the work.}
\address{\affa}
\author{Si-Yun Zhou}
\thanks{These authors contributed equally to the work.}
\address{\affb}
\address{\affc}
\author{Yun-Hao Shi}
\thanks{These authors contributed equally to the work.}
\address{\affb}
\author{Kaixuan Huang}
\thanks{These authors contributed equally to the work.}
\address{\affa}
\author{Z. H. Yang}
\address{\affb}
\address{\affc}
\author{Jingning Zhang}
\email{zhangjn@baqis.ac.cn}
\address{\affa}
\author{\mbox{Kui Zhao}}
\address{\affa}
\author{Yueshan Xu}
\address{\affa}
\author{Hao Li}
\address{\affa}
\author{S. K. Zhao}
\address{\affa}
\author{Yulong Feng}
\address{\affa}
\author{Guangming Xue}
\address{\affa}
\author{Yu Liu}
\address{\affb}
\address{\affc}
\author{Wei-Guo Ma}
\address{\affb}
\address{\affc}
\author{\mbox{Cai-Ping Fang}}
\address{\affb}
\address{\affc}
\author{Hao-Tian Liu}
\address{\affb}
\address{\affc}
\author{Yong-Yi Wang}
\address{\affb}
\address{\affc}
\author{Kai Xu}
\address{\affa}
\address{\affb}
\address{\affd}
\author{Haifeng Yu}
\email{hfyu@baqis.ac.cn}
\address{\affa}
\address{\affd}
\author{Heng Fan}
\email{hfan@iphy.ac.cn}
\address{\affa}
\address{\affb}
\address{\affd}
\author{S. P. Zhao}
\email{spzhao@iphy.ac.cn}
\address{\affa}
\address{\affb}

\pacs{xxx}

\maketitle

\clearpage
\newpage
\appendix
\onecolumngrid
	
\section{DERIVATION OF THE EFFECTIVE HAMILTONIAN}

\vspace{5mm}

We consider a chain of frequency-tunable transmon qubits (below we use `qubit' and `qutrit' interchangeably for convenience). The time-dependent many-body Hamiltonian reads
\begin{equation}
    \hat H(t)=\sum_{l=1}^{L}\left[\omega_l(t)\hat a_l^\dagger\hat a_l+\frac{\alpha_l}{2}\hat a_l^\dagger\hat a_l^\dagger\hat a_l\hat a_l\right]+\sum_{l=1}^{L-1}g_l\left(\hat a_l^\dagger\hat a_{l+1}+\hat a_{l+1}^\dagger\hat a_l\right),
\end{equation}
here $\hat a_l^\dagger$ ($\hat a_l$) is the bosonic creation (annihilation) operator. The frequency of each transmon qubit is periodically modulated by
\begin{equation}
    \omega_l(t)=\omega_{\rm res}+\Omega_l\cos\left(\nu_l t+\phi_l \right)
    \label{eq:omega}
\end{equation}
with $\Omega_l$, $\nu_l$, and $\phi_l$ being the amplitude, frequency, and phase of the periodic modulation, respectively. Usually, the negative anharmonicity ($\alpha_l<0$) and the weak tunneling strength $(\left|g_l/\alpha_l\right|\ll 1$) place the native Hamiltonian in the strong-interaction regime, where it can be approximated by a hard-core boson model.

Let us define the unitary transformation
\begin{equation}
\hat U_0(t)={\mathcal T}\exp\left[-i\int_0^t\hat H_0(t)dt\right]
\end{equation}
with 
\begin{equation}
 \label{eq:H_unitary_transform}
    \hat H_0(t)=\sum_{l=1}^{L}\left[\omega_l(t)\hat a_l^\dagger\hat a_l+\frac{\alpha_l}{2}\hat a_l^\dagger\hat a_l^\dagger\hat a_l\hat a_l\right],
\end{equation}
where $\mathcal{T}$ denotes time-ordering. The transformed Hamiltonian becomes 
\begin{eqnarray}
\hat H'&=&\hat U^\dagger_0(t)\left(\hat H(t)-\hat H_0(t)\right)\hat U_0(t)\\
&=&\sum_{l=1}^{L-1}g_l\left(\hat U_0^\dagger(t)\hat a_l^\dagger\hat a_{l+1}\hat U_0(t)+{\rm h.c.}\right).
\end{eqnarray}
Both $\hat{H}_0(t)$ and $\hat{U}_0(t)$ are diagonal in the Fock basis, namely
\begin{eqnarray}
    \hat H_0(t)&=&\sum_{l=1}^L\sum_{n_l}\left[n_l\omega_l(t)+\frac{\alpha_l}{2} n_l\left( n_l -1\right)\right]\ketbra{n_l}{n_l}, \\
    \hat U_0(t)&=&\prod_{l=1}^L\left(\sum_{n_l}\exp\left(-i\int_0^t\left(n_l\omega_l(t)+\frac{\alpha_l}{2}n_l\left(n_l-1\right)\right)dt\right)\ketbra{n_l}{n_l}\right).
\end{eqnarray}
To simplify the notations, we set $\omega_{l,0}=\omega_0$ and $\phi_l=0$ and consider that only the qubits on the even sites are driven, such that $\Omega_l=\Omega$, $\nu_l=\nu$ if $l$ is even, and $\Omega_l=0$ other wise.
In the subspace of up to 2 excitations in the whole chain, the relevant transition terms in the effective Hamiltonian are
\begin{eqnarray}
\hat U_0^\dagger\ket{01}\bra{10}_{l,l+1}\hat U_0&=&\exp\left(\pm\frac{i\Omega}{\nu}\sin\nu t\right)\ket{01}\bra{10}_{l,l+1},\\
\hat U_0^\dagger\ket{11}\bra{20}_{l,l+1}\hat U_0&=&\exp\left(-i\alpha_lt\pm\frac{i\Omega}{\nu}\sin\nu t\right)\ket{11}\bra{20}_{l,l+1},\\
\hat U_0^\dagger\ket{11}\bra{02}_{l,l+1}\hat U_0&=&\exp\left(-i\alpha_lt\pm\frac{i\Omega}{\nu}\sin\nu t\right)\ket{11}\bra{02}_{l,l+1},
\end{eqnarray}
where the positive or negative sign depends on $l$ being odd or even. 

Using the Jacobi-Anger expansion
\begin{eqnarray}
\exp\left(iz\cos\theta\right)&=&\sum_{m=-\infty}^\infty i^mJ_m(z)e^{im\theta},\\
\exp\left(\pm\frac{i\Omega}{\nu}\sin\nu t\right)&=&\ldots \mp J_{-1}\left(\frac{\Omega}{\nu}\right)e^{i\nu t}+J_0\left(\frac{\Omega}{\nu}\right)\mp J_{1}\left(\frac{\Omega}{\nu}\right)e^{-i\nu t}+\ldots,
\end{eqnarray}
and keeping only the near-resonant terms under the condition that $\alpha_l,\nu\gg g_l$, we obtain
\begin{eqnarray}
\hat U_0^\dagger g_l\ket{01}\bra{10}_{l,l+1}\hat U_0&=&g_lJ_0\left(\frac{\Omega_l}{\nu_l}\right)\ket{01}\bra{10}_{l,l+1}+{\mathcal O}\left(g_l^2\right),\\
\hat U_0^\dagger \sqrt{2}g_l\ket{11}\bra{20}_{l,l+1}\hat U_0&=&\sqrt{2}g_lJ_1\left(\frac{\Omega}{\nu}\right)e^{-i(\alpha_l+\nu)t}\ket{11}\bra{20}_{l,l+1}+{\mathcal O}\left(g_l^2\right),\\
\hat U_0^\dagger \sqrt{2}g_l\ket{11}\bra{02}_{l,l+1}\hat U_0&=&\sqrt{2}g_lJ_1\left(\frac{\Omega}{\nu}\right)e^{-i(\alpha_l+\nu)t}\ket{11}\bra{02}_{l,l+1}+{\mathcal O}\left(g_l^2\right),
\end{eqnarray}
where the magnitude of $\alpha_l+\nu$ is assumed to be comparable to the tunneling strength $g$. 

In order to obtain a time-independent effective Hamiltonian, we introduce a second rotating frame defined by 
\begin{equation}
\hat U'(t)=\prod_{l}\left[\ketbra{0}{0}_l+\ketbra{1}{1}_l+e^{-i(\alpha_l+\nu)t}\ketbra{2}{2}_l\right]~. 
\end{equation}
In this frame, the effective Hamiltonian becomes
\begin{eqnarray}
\hat H_{\rm eff}&=&\sum_{l=1}^L\left(\alpha_l+\nu\right)\ketbra{2}{2}_l
+\sum_{l=1}^{L-1}g_l\left(J_0\left(\varepsilon\right)\ket{01}\bra{10}_{l,l+1}+\sqrt{2}J_1\left(\varepsilon\right)\ket{11}\bra{20}_{l,l+1}\right. \nonumber \\
&&\left.+\sqrt{2}J_1\left(\varepsilon\right)\ket{11}\bra{02}_{l,l+1}\right)
\end{eqnarray}
with $\varepsilon\equiv \Omega/\nu$. By requiring $J_0\left(\varepsilon^*\right)=J_1\left(\varepsilon^*\right)$ or $\varepsilon^*\approx{1.4347}$, we can rewrite the above effective Hamiltonian in terms of creation and annihilation operators,
\begin{equation}
    \hat H_{\rm eff}=\sum_l g_{l,{\rm eff}}\left(\hat a_l^\dagger\hat a_{l+1}+\hat a^\dagger_{l+1}\hat a_l\right)-\sum_l\frac{U_{l,\rm eff}}{2}\hat a_l^\dagger\hat a_l^\dagger\hat a_l\hat a_l, 
\end{equation}
with
\begin{eqnarray}
U_{l,{\rm eff}}=|\alpha_l+\nu|,\quad g_{l,{\rm eff}}=g_lJ_0\left(\varepsilon^*\right).
\label{eq:Ueffgeff}
\end{eqnarray}
Notably, the same results can be achieved when neighboring qubits are driven by longitudinal fields with a phase difference of $\phi = \phi_{l+1}-\phi_l $. 

In the main text, we denote $U_{l,{\rm eff}}$ and $g_{l,{\rm eff}}$ as $U$ and $J$ commonly used for the Bose-Hubbard model, and we emphasize that the above effective Hamiltonian is only valid in the subspace of up to 2 excitations in the whole chain.

\vspace{8mm}
\section{THE IMPACT OF COUPLERS IN FLOQUET ENGINEERING}
\vspace{5mm}

Given that the qubits in our processor are interconnected through couplers, a key aspect of this study involves the analysis of how the couplers influence the system's dynamical evolution. For simplicity, we consider two transmon qubits mediated by a frequency-tunable coupler. The total time-dependent Hamiltonian is
\begin{eqnarray}
	\hat H(t)&=&\sum_{i=1,2}\left(\omega_i\hat a_i^\dag\hat a_i+\frac{\alpha_i}{2}\hat a_i^\dag\hat a_i^\dag\hat a_i\hat a_i\right)+\left(\omega_c\hat a_c^\dag\hat a_c+\frac{\alpha_c}{2}\hat a_c^\dag\hat a_c^\dag\hat a_c\hat a_c\right)\nonumber\\
	&&+\sum_{i=1,2}g_{i,c}\left(\hat a_i^\dag\hat a_c+\hat a_c^\dag\hat a_i\right)+g_{1,2}\left(\hat a_1^\dag\hat a_2+\hat a_2^\dag\hat a_1\right),
\end{eqnarray}
with $\hat a_c$ ($\hat a_c$) being the bosonic creation (annihilation) operator for the coupler. With the Schrieffer-Wolff transformation described by the following unitary operator~\cite{PhysRevApplied.10.054062}
\begin{eqnarray}
	\hat U=\exp\left[\sum_{i=1,2}\frac{g_{i,c}}{\omega_i-\omega_c}\left(\hat a_i^\dagger\hat a_c-\hat a_c^\dagger\hat a_i\right)\right],
\end{eqnarray}
we are able to decouple the coupler degrees of freedom and obtain the effective Hamiltonian that only involves the transmon qubits:
\begin{eqnarray}
	\hat H_{\rm eff}&=&\hat U\hat H\hat U^\dag\nonumber\\
	&=&\sum_{i=1,2}\left(\tilde\omega_i\hat a_i^\dagger\hat a_i+\frac{\tilde\alpha}{2}\hat a_i^\dagger\hat a_i^\dagger\hat a_i\hat a_i\right)+\tilde g_{1,2}\left(\hat a_1^\dagger\hat a_2+\hat a_2^\dagger\hat a_1\right)
\end{eqnarray}
with the dressed parameters of
\begin{eqnarray}
	\tilde\omega_i&=&\omega_i+\frac{g_{i,c}^2}{\Delta_i},\quad \tilde\alpha_i=\alpha_i,\\
	\tilde g_{1,2}&=&\frac{g_{1,c}g_{2,c}}{2}\left(\frac{1}{\Delta_1}+\frac{1}{\Delta_2}\right)+g_{1,2},
\end{eqnarray}
where $\Delta_i=\omega_i-\omega_c$ is the qubit-coupler detuning.

As discussed above, we consider the situation in which the frequency of the second qubit is modulated by a time-dependent sinusoidal external flux, with the central frequency in resonance with that of the first qubit. That is to say, the raw frequencies of the two qubits are 
\begin{eqnarray}
	\omega_1=\omega_{\rm res},\quad \omega_2(t)=\omega_{\rm res}+\Omega\cos\nu t,
\end{eqnarray}
here $\nu$ and $\Omega$ are again the frequency and amplitude of the driving flux, respectively. Following the above procedure, we arrive at a time-dependent effective Hamiltonian by substituting $\tilde \omega_2$ and $\tilde g_{1,2}$ with the following time-dependent expressions:
\begin{eqnarray}
	\tilde\omega_2(t)&=&\omega_{\rm res}+\frac{g_{2,c}^2}{\Delta_{\rm res}}+\Omega\left(1-\frac{g_{2,c}^2}{\Delta_{\rm res}^2}\right)\cos\nu t, \label{2qa}\\
	\tilde g_{1,2}(t)&=&\frac{g_{1,c}g_{2,c}}{\Delta_{\rm res}}\left(1-\frac{\Omega}{2\Delta_{\rm res}}\cos\nu t\right)+g_{1,2}, \label{2qb}
\end{eqnarray}
with $\Delta_{\rm res}=\omega_{\rm res}-\omega_c$. From Eqs.~(\ref{2qa}, \ref{2qb}), we see that the coupler does not alter the frequency of the longitudinal driving, but it introduces an additional oscillating term in the effective qubit-qubit coupling, which can be diminished by increasing the frequency detuning between the resonant qubits and the coupler. Meanwhile, the finite coherence times of the qubits limit the available evolution time, and thus require the BH simulator to operate with relatively strong coupling. Balancing these two considerations, we set the frequency detuning between the resonant qubits and the coupler to $\Delta_{\rm res}/2\pi=1.2$~GHz. Under this experimental setting, we measured the dependence of the effective qubit–qubit coupling on the longitudinal field strength. The experimental data, as presented in Fig.~\ref{fig:zpa_compensation}(c) below, are in excellent agreement with the theoretical prediction proportional to the Bessel function $J_0\left(\Omega/\nu\right)$.
 
\vspace{8mm}
\section{DEVICE INFORMATION}
\vspace{5mm}

The superconducting processor employed in this work comprises 9 transmon qutrits and 12 couplers fabricated via flip-chip technology with a tantalum base layer~\cite{li2021vacuum, fab2}, arranged in a $3\times3$ square lattice (schematically shown in Fig.~1 in the main text). In the experiment, we decouple qutrits Q$_1$ through Q$_8$ from other components by setting the frequencies of related couplers to $\sim$~3.2~GHz in decoupling state, and form one-dimensional chains with uniform coupling strength of $g=$11~MHz. The detailed device parameters are listed in Table~\ref{tab:table1}. 

\begin{table}
\caption{\label{tab:table1}
	\small{\textbf{Device parameters for qutrits Q$_1$-Q$_8$ used in this work.} $\omega_r$, $\omega_{\rm{max}}$, $\omega_{\rm{idle}}$, and $\omega_{\rm{res}}$ denote the readout resonator frequency, maximum qubit frequency, qubit idle frequency, and resonant frequency, respectively. $\alpha_{\rm idle}$ and $\alpha_{\rm res}$ represent the qutrit anharmonicity at the idle frequency and resonant frequency, respectively. $T_{1,ge}$ and $T_{1,ef}$ are the relaxation times for $|e\rangle \rightarrow |g\rangle$ and $|f\rangle \rightarrow |e\rangle$ transitions, and $T_{2,ge}$ and $T_{2,ef}$ are the corresponding dephasing times at the idle point, where $|g\rangle$, $|e\rangle$, and $|f\rangle$ are the qutrit ground, first-excited, and second-excited states, respectively. $F_g$, $F_e$, and $F_f$ indicate 8-qutrit simultaneous readout fidelities for the corresponding states. Single-qubit gate fidelities in the last row are calibrated via randomized benchmarking (RB).}}
    \begin{tabular}{c|cccccccc}
    		\hline\hline
    		Qubits & $\rm Q_1$ & $\rm Q_2$ & $\rm Q_3$ & $\rm Q_4$ & $\rm Q_5$ & $\rm Q_6$& $\rm Q_7$& $\rm Q_8$\\ 
      \hline
      $\omega_r$ (GHz)& 7.346	&7.352	&7.356	&7.315&	7.302&	7.245&	7.285	&7.332 \\
      $\omega_{\rm max}$ (GHz) & 5.058&	5.234&	5.119&	5.149&	5.076&	5.071&	4.986&	5.065\\
      $\omega_{\rm idle}$ (GHz) &5.058 & 4.484	&5.106 &	4.381&	5.038&	4.572&	4.986&	4.434\\
      $\alpha_{\rm idle}$ (MHz)&-192.8&	-211.1&	-196.6&	-210.8&	-197.7&	-210.1&	-197.3&	-209.9\\
      
      $T_{1,ge}$ ($\mu$s)&22.26&	30.31&	18.80&	46.34&	24.24&	21.79&	21.24&	60.76\\
      $T_{1,ef}$ ($\mu$s)&28.48&	21.43&	40.66&	76.403&	20.59&	33.51&	32.29&	27.55\\
      $T^*_{2,ge}$($\mu$s)&10.87&	0.84&	6.48&	0.97&	2.53&	1.17&	14.92&	1.24\\
      $T^*_{2,ef}$($\mu$s)&10.39&	0.69&	4.29&	0.91&	2.54&	1.39&	5.68&	1.09\\
      $F_g$($\%$)&95.5 &94.4&	94.0&	97.5&	96.4&	96.7&	90.9&	95.0	\\
      $F_e$($\%$) & 84.2 &91.1&	90.1&	94.8&	95.0&	94.9&	84.9&	86.1\\
      $F_f$($\%$) & 72.3 & 85.2&	76.6&	89.4&	88.2&	92.1&	82.5&	79.3	\\
      $\omega_{\rm res}$ (GHz)& 4.5& 4.5& 4.5& 4.5& 4.5& 4.5& 4.5& 4.5\\
      $\alpha_{\rm res}$ (MHz)& -207.5&-210.9& -200.8& -207.5&   -200.9& -210.4& -200.8& -209.1\\
     1Q RB ($\%$)& 99.971& 99.849& 99.943& 99.906& 99.959& 99.886& 99.941& 99.889\\
  \hline\hline
	\end{tabular}
\vspace{5mm}
\end{table}

The parameter $\alpha_{\rm res}$ in Table~\ref{tab:table1} is denoted as $\alpha_{l}$ in the main text. We can see that it ranges approximately from $−$201 to $−$211 MHz. Since the on-site interaction is given by $U=\left|\nu+\alpha_l\right|$ while $\nu$ is uniform, we define $U=\left|\nu+\bar{\alpha}_l\right|$ in this work where $\bar{\alpha}_l$ is the averaged value of $\alpha_{\rm res}$ given in Table~\ref{tab:table1}. Refer to Ref.~\cite{wang_demonstration_2024} for additional information about the measurement setup, device parameters, and single-qutrit calibration procedures.

\vspace{8mm}
\section{QUTRIT READOUT}
\vspace{5mm}

\begin{figure}[t]
	\centering
	\includegraphics[width=0.7\linewidth]{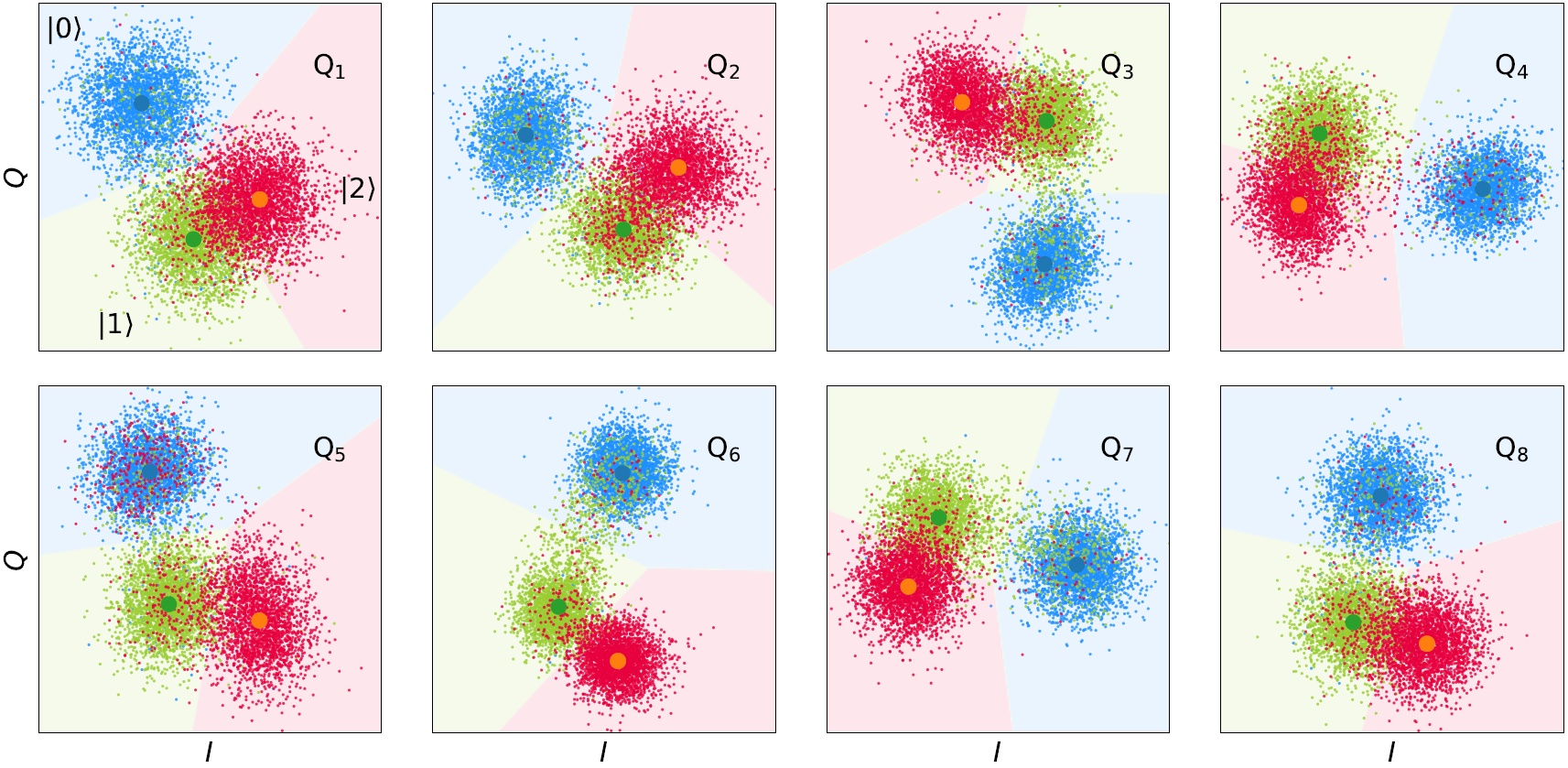}
	\caption{I-Q plots of simultaneous signal-shot measurements of Q$_1$ through Q$_8$.}
	\label{fig:readoutIQ}
\end{figure}

The present experiment requires reading out the qutrit three states: the ground state $|0\rangle$ (or $|g\rangle$ state), the first-excited state $|1\rangle$ (or $|e\rangle$ state), and the second-excited state $|2\rangle$ (or $|f\rangle$ state). By adjusting the readout frequencies, readout powers, readout durations, qubit frequencies, and the working points of the Josephson parametric amplifiers, high-fidelities for simultaneous readout of each state and each qutrit are achieved, as can be seen in Table~\ref{tab:table1}.  Figure~\ref{fig:readoutIQ}  shows the simultaneously measured I-Q data for 3000 state discriminations of the 8 qutrits from Q$_1$ to Q$_8$.

\vspace{8mm}
\section{AMPLITUDE MODULATION IN FLOQUET ENGINEERING}
\vspace{5mm}

To tune the qutrit on-site interaction $U$ in our experiments, we bias even-labeled qutrits with ac magnetic flux that periodically modulates driven qutrit frequency according to Eq.~\ref{eq:omega}. For frequency-tunable transmon qubits with symmetric Josephson junctions, the flux-frequency relation is given by~\cite{PhysRevA.76.042319}:
\begin{equation}
    V_z = \frac{1}{k}\arccos\left[\frac{\omega_{\rm res} + \Omega\sin(\nu t + \phi) + E_C}{8E_JE_C}\right] - \frac{b}{k}
\end{equation}
where $\omega_{\rm res}$ is the qubit average frequency, $kV_z+b = \pi \Phi_e/\Phi_0$ is the ratio of the external flux $\Phi_e$ to the flux quantum $\Phi_0$, $E_J$ denotes the Josephson energy without flux in the loop of the superconducting quantum interference device (SQUID), and $E_C$ is the charging energy. From the target $U$, we derive the frequency and amplitude of longitudinal field to construct the actual Z-control waveform.

\begin{figure}[b]
    \centering
    \includegraphics[width=0.95\linewidth]{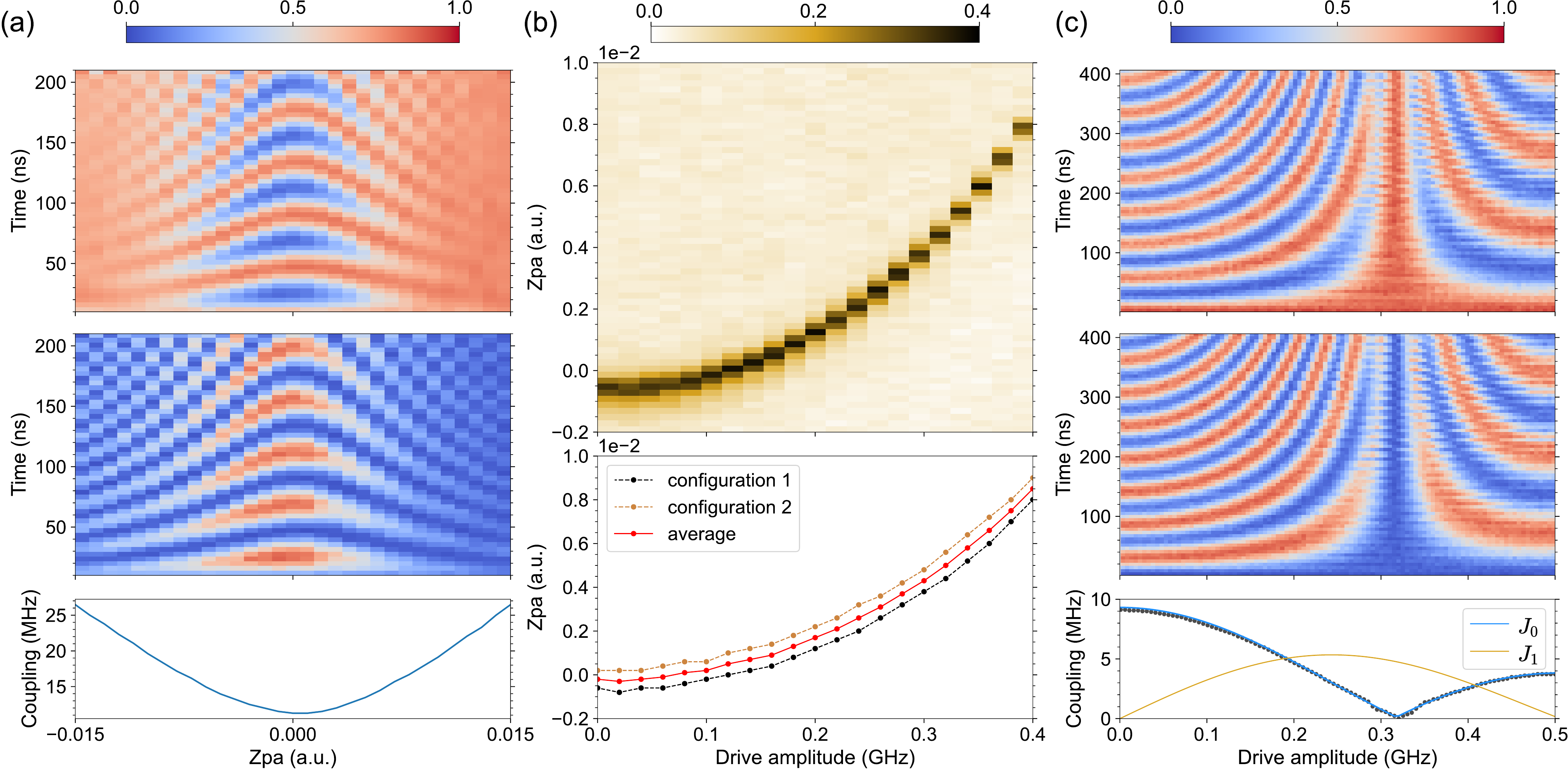}
    \caption{\small{\textbf{Calibration of longitudinal drive field.}
    (a) Coupling strength calibration between two adjacent qubits via vacuum Rabi oscillations in the $|01\rangle$–$|10\rangle$ subspace without longitudinal drive. (b) Spectral characterization of drive-induced qubit frequency shifts. (c) Functional dependence of effective qubit-qubit coupling strength on longitudinal drive amplitude calibrated through vacuum Rabi oscillations. $J_0$ and $J_1$ indicate the zeroth- and first-order Bessel functions, respectively.}}
    \label{fig:zpa_compensation}
\end{figure}

The coupling strength in the original Hamiltonian is fixed at $g=$~11~MHz calibrated by performing two qubit swap experiments as shown in Fig.~\ref{fig:zpa_compensation}(a).
The amplitude-to-frequency ratio of the longitudinal drive field is theoretically a constant value of $\sim$ 1.4347. However, a scaling factor exists between the experimentally applied drive field and theoretical values, determined from the field amplitude versus coupling strength relationship. We first correct qubit frequency shifts induced by the longitudinal drive and fit their relationship with a polynomial function [see Fig.~\ref{fig:zpa_compensation}(b)]. Subsequently, vacuum Rabi oscillations are measured with varying drive strengths. The scaling factor is extracted through zeroth-order Bessel function fitting, and the first crossover point between zeroth- and first-order Bessel functions determines the drive amplitude $\Omega$ on individual qubits, as shown in Fig.~\ref{fig:zpa_compensation}(c).

\vspace{8mm}
\section{ENERGY SPECTRUM MEASUREMENT}
\vspace{5mm}

\subsection{Dynamical phase calibration}

\begin{figure}[t]
    \centering
    \includegraphics[width=0.95\linewidth]{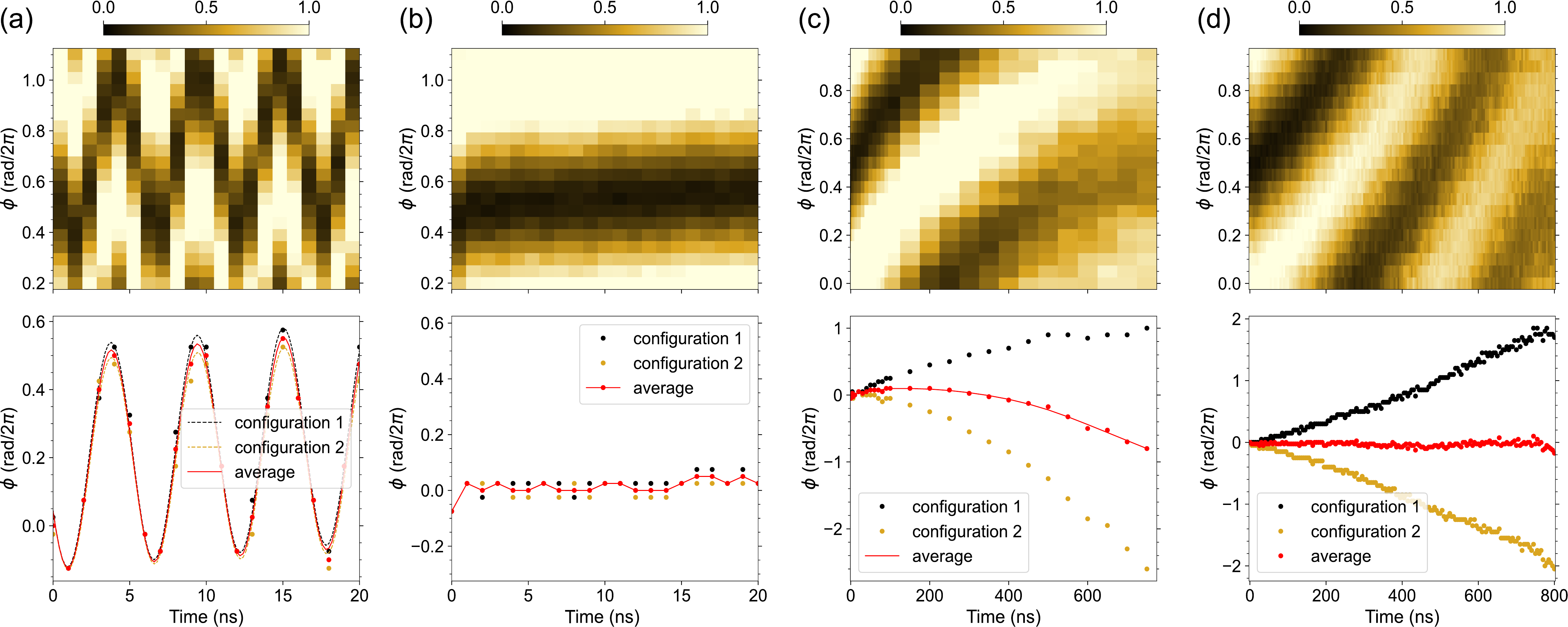}
    \caption{\textbf{Phase calibration.} The experimental results of the $|1\rangle$-state probability evolution before (a) and after (b) the phase calibrated with sinusoidal form. The $|1\rangle$-state probability evolution before (c) and after (d) the phase calibrated with interpolation.}
    \label{fig:phase_calibration}
\end{figure}

In our experiments, we tune all qubits to a common resonant point from idle points by rectangular Z pulses. The total dynamical phase can be split into three parts. The first one comes from the difference of the qubit frequencies between the idle points $\omega_{\rm idle}$ and the resonant point $\omega_{\rm res}$:
\begin{equation}
    \Delta\varphi_1 = (\omega_{\rm res} - \omega_{\rm idle})t,
\end{equation}
where $t$ is the evolution time. This part can be add to the circuit in Fig.~1(c) in the main text as a virtual-Z gate before measurement. The second one is caused by the control deviation, i.e., the driving amplitude felt by the driven qubit deviates from the output of the control system, shown in Fig. \ref{fig:phase_calibration}(a). This part complies with sinusoidal form:
\begin{equation}
    \Delta \varphi_2 = a\cos(\nu t +b),
\end{equation}
and as shown in Fig. \ref{fig:phase_calibration}(b), it can be almost completely compensated.  The final part results from the imperfect rectangle Z pulse, including the rising and falling edges, distortions, etc. [see Fig.~\ref{fig:phase_calibration}(c)], which can be expressed as
\begin{equation}
    \Delta\varphi_3 = \int_o^t [\omega_{\rm res}-\omega_{\rm actual}(t)]dt.
\end{equation}
Here $\omega_{\rm actual}(t)$ represents the actual frequency on the driven qubit. By adding this phase accumulation directly to the circuit,  the accumulated extra dynamical phase can be removed, as shown in Fig. \ref{fig:phase_calibration}(d).

\subsection{Time-dependent correlation function}

To obtain the time-dependent correlation function for the energy spectrum measurement~\cite{Roushan2017, shi2023, xiang_simulating_2023}, we consider the initial state prepared in the form of
\begin{equation}
    |\Psi_0\rangle_{(m,n)} = |0\rangle_1 \otimes |0\rangle_2\otimes \dots \otimes \frac{1}{\sqrt{2}} \left(|0\rangle_m + |1\rangle_m\right)\otimes \dots \otimes \frac{1}{\sqrt{2}} \left(|0\rangle_n + |1\rangle_n\right)\otimes \dots \otimes|0\rangle_{N-1}\otimes|0\rangle_N\;(m\neq n).
\end{equation}
The state at time $t$ expanded with eigenstates is 
\begin{equation}
    |\Psi_t\rangle_{(m,n)}=e^{-iHt}|\Psi_0\rangle_{(m,n)}=\frac{1}{2}\left( |Vac\rangle + \sum_\beta C_{\beta,(m,n)} e^{-iE_\beta^{(2)}t}|\phi_\beta^{(2)}\rangle \right) + \frac{1}{2} \left( \sum_\alpha (C_{\alpha,m} + C_{\alpha,n}) e^{-iE_\alpha^{(1)}t}|\phi_\alpha^{(1)}\rangle \right),
\end{equation}
where $\beta\in\{1,2,\dots,\frac{1}{2}N(N+1)\}$, $|\phi_\beta^{(2)}\rangle$ is an energy eigenstate in the two-photon manifold  with the corresponding energy $E_\beta^{(2)}$ and $C_{\beta,(m,n)}=\langle \phi_\beta^{(2)}|\mathbf{1}_m,\mathbf{1}_n\rangle$. Since the second excited states are considered in our experiments, the amplitude of two-point correlation is constructed as

\begin{align}
     \chi_{mn} \equiv \langle\hat{a}_m \hat{a}_n\rangle 
    &= \langle \sigma_{m,01}^{x}\sigma_{n,01}^{x}\rangle + 
      i\langle \sigma_{m,01}^{x}\sigma_{n,01}^{y} \rangle + 
      i \langle \sigma_{m,01}^{y}\sigma_{n,01}^{x} \rangle - 
      \langle \sigma_{m,01}^{y}\sigma_{n,01}^{y} \rangle \nonumber \\
    &+ \sqrt{2} \langle \sigma_{m,01}^{x}\sigma_{n,12}^{x} \rangle + 
       \sqrt{2}i \langle \sigma_{m,01}^{x}\sigma_{n,12}^{y}) \rangle + 
       \sqrt{2}i \langle \sigma_{m,01}^{y}\sigma_{n,12}^{x} \rangle -
       \sqrt{2} \langle \sigma_{m,01}^{y}\sigma_{n,12}^{y} \rangle \nonumber \\
    &+ \sqrt{2} \langle \sigma_{m,12}^{x}\sigma_{n,01}^{x} \rangle + 
       \sqrt{2}i \langle \sigma_{m,12}^{x}\sigma_{n,01}^{y} \rangle + 
       \sqrt{2}i \langle \sigma_{m,12}^{y}\sigma_{n,01}^{x} \rangle -
       \sqrt{2} \langle \sigma_{m,12}^{y}\sigma_{n,01}^{y} \rangle \nonumber \\
    &+ 2 \langle \sigma_{m,12}^{x}\sigma_{n,12}^{x} \rangle + 
       2i \langle \sigma_{m,12}^{x}\sigma_{n,12}^{y} \rangle +
       2i \langle \sigma_{m,12}^{y}\sigma_{n,12}^{x} \rangle -
       2 \langle \sigma_{m,12}^{y}\sigma_{n,12}^{y} \rangle,
    \label{eq:chi_t}
\end{align}
in which the subscript 01 (12) means the measurement between energy level 0 (1) and 1 (2), and superscript x (y) means the measurement is Pauli-X (Pauli-Y). Its expectation value takes the form
\begin{equation}
    \chi_{mn}(t) = \frac{1}{4}\sum_\beta |C_{\beta,(m,n)}|^2 e^{-iE_\beta^{(2)}t},
\end{equation}
which project out the one-photon component to avoid measuring the energy differences $E_\beta^{(2)}-E_\alpha^{(1)}$.

There are 16 expectation terms in Eq.~\ref{eq:chi_t}. To simplify the measurements, we simulate the system's dynamics and find that only the first four items have the information of the energy levels. Therefore we only measure $\langle \sigma_{m,01}^{x}\sigma_{n,01}^{x}\rangle,\langle \sigma_{m,01}^{x}\sigma_{n,01}^{y} \rangle,\langle \sigma_{m,01}^{y}\sigma_{n,01}^{x} \rangle,\langle \sigma_{m,01}^{y}\sigma_{n,01}^{y} \rangle$ to calculate $\chi_{mn}(t)$.

\begin{figure}[t]
	\centering
	\includegraphics[width=0.95\textwidth]{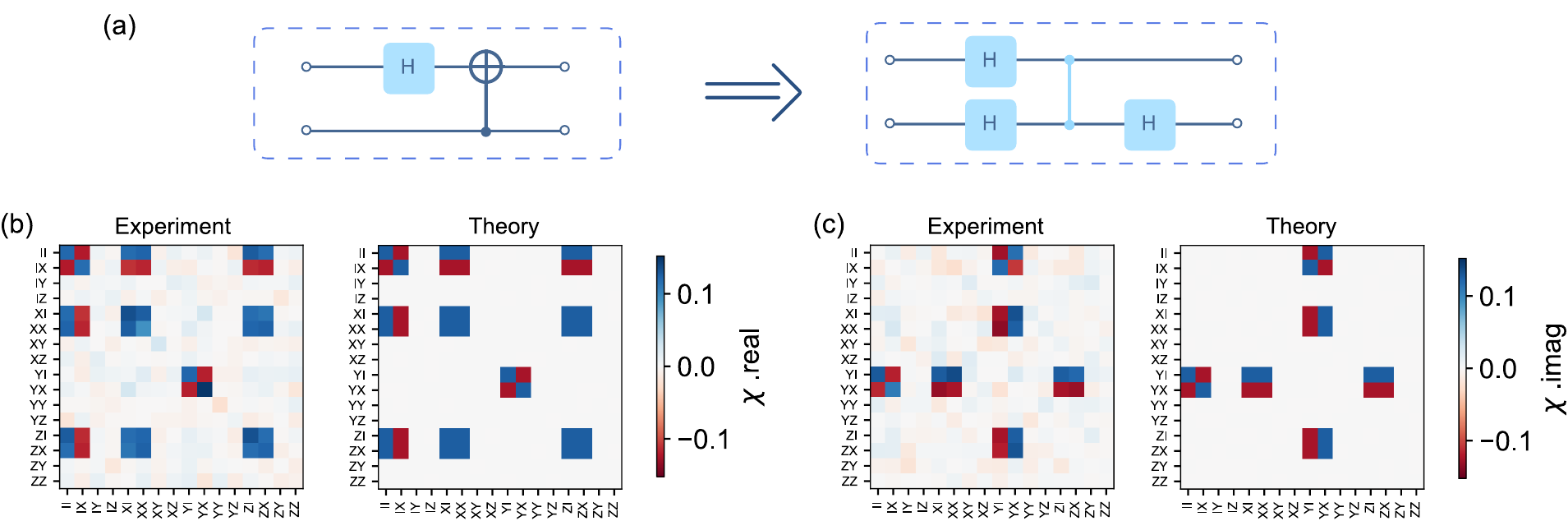}
	\caption{\small{\textbf{Quantum process tomography for a Bell state $|\psi\rangle = (|00\rangle +|11\rangle)/\sqrt{2}$.} (a) The gate sequence to prepare the Bell state. (b) and (c) The real (b) and imaginary (c) parts of the experimental and theoretical results of the Bell state. The corresponding fidelity is 98.28$\%$.}}
	\label{fig:bellexample}
\end{figure}

\vspace{8mm}
\section{BELL STATE PREPARATION}
\vspace{5mm}

To prepare the Bell state $|\psi\rangle = (|00\rangle + |11\rangle)/\sqrt{2}$ between two qubits, we execute the gate sequence shown in Fig.~\ref{fig:bellexample}(a). Here, we utilize a CZ gate and a Hadamard gate instead of a CNOT gate. The conditional phases of the CZ gate are implemented using microwave gates, specifically $R_z(\theta) = R_y(-\pi/2) R_x(\theta) R_y(\pi/2)$. Figures~\ref{fig:bellexample}(b) and (c) display the $\chi$ matrix measured via quantum process tomography, with a corresponding fidelity of 97.37$\%$.

\vspace{8mm}
\section{QUANTUM STATE TOMOGRAPHY OF THE QUTRIT SYSTEM} 
\vspace{5mm}

Quantum state tomography (QST) is a fundamental technique for fully reconstructing the density matrix of a quantum state generated from a given input. The reconstructed density matrix enables calculation of key quantum information quantities, including negativity and quantum discord. The superconducting platform possesses manageable and favorable complexity scaling for quantum state tomography in multi-level systems. Specifically, for one (two) qutrit, quantum state tomography requires 6 ($6^2$) measurement operations to reconstruct the density matrix, while for one (two) qudit, it requires 12 ($12^2$) measurement operations, as demonstrated in the literature~\cite{liu_performing_2023}\cite{vbh4-lysv}. So by increasing the number of measurement operations accordingly, the approach of quantum state tomography can be extended to reconstruct the density matrices of qudits and other multi-level systems.

\begin{figure}[t]
	\centering
	\includegraphics[width=0.8\textwidth]{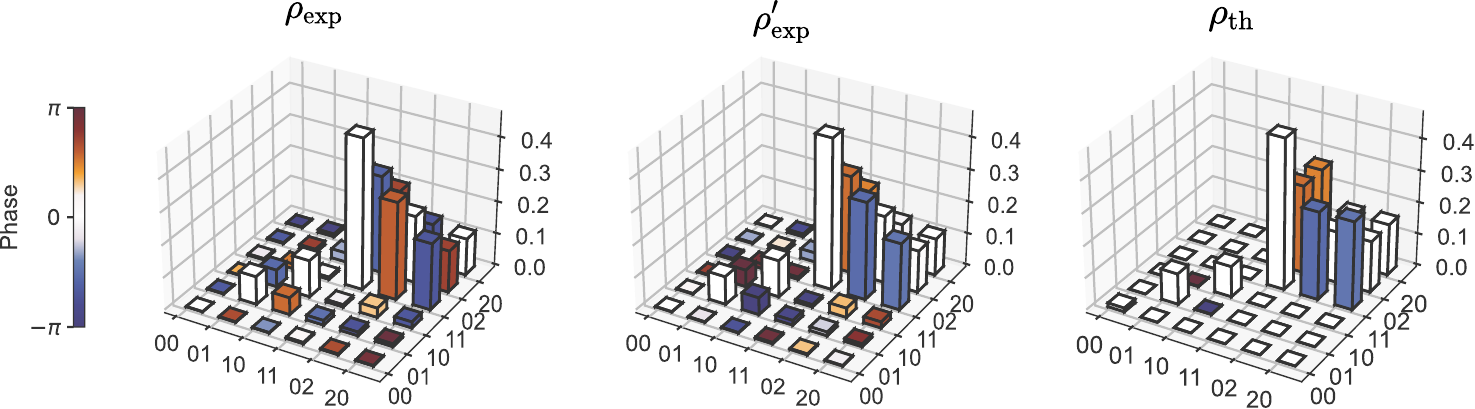}
	\caption{\small{\textbf{The histogram representation of the  two-qutrit density matrices.} The leftmost subplot corresponds to the experimentally measured $\rho_{\rm exp}$, with a fidelity of $81.3\%$ to the theoretical $\rho_{\rm th}$ illustrated in the rightmost subplot; the middle subplot displays the result of $\rho'_{\rm exp}$ by applying a single-qubit phase gate numerically to the experimental $\rho_{\rm exp}$, yielding a fidelity of $96.5\%$.}}
	\label{fig:qutrit_density_histogram}
\end{figure}

For the qutrit state reconstruction, we adopt the methodology from Refs.~\cite{liu_performing_2023} and \cite{ringbauer_universal_2022}. Specifically, we select six unitary operators $\hat{U}_l$ as tomography gates:
\begin{align}
    \hat{U}_0 &= \hat{R}_{0,1}(\tfrac{\pi}{2},0),\\
    \hat{U}_1 &= \hat{R}_{0,1}(-\tfrac{\pi}{2},\tfrac{\pi}{2}),\\
    \hat{U}_2 &= \hat{R}_{1,2}(\tfrac{\pi}{2},0),\\
    \hat{U}_3 &= \hat{R}_{1,2}(-\tfrac{\pi}{2},\tfrac{\pi}{2}),\\
    \hat{U}_4 &= \hat{R}_{1,2}(\tfrac{\pi}{2},0)\hat{R}_{0,1}(\tfrac{\pi}{2},0),\\
    \hat{U}_5 &= \hat{R}_{1,2}(-\tfrac{\pi}{2},\tfrac{\pi}{2})\hat{R}_{0,1}(\tfrac{\pi}{2},0).
\end{align}
For any multi-qutrit system, the unitary rotation set comprises all possible combinations of operators $\hat{U}_k$ ($k=0,\dots,5$) applied to individual qutrits. 

After applying a unitary rotation $\hat{U}_i$ to an unknown quantum state $\hat{\rho}$, the projection probabilities onto eigenstates $|k\rangle$ are given by:
\begin{equation}
    P_{i,k} = \langle k | \hat{U}^{\dagger}_i \hat{\rho} \hat{U}_i | k \rangle.
\end{equation}
Measurements of $P_{i,k}$ yield a system of $6^N$ equations, where $N$ denotes the number of qutrits. The density matrix $\hat{\rho}$ is then reconstructed using standard maximum-likelihood estimation (MLE). 

In the present experiment, we need to reconstruct the joint density matrix of a two-qutrit system. Employing a complete set of 36 orthogonal measurement bases $\otimes^2 \{\hat{U}_i \mid i = 1,\dots,6\}$, quantum state tomography produces a $9 \times 9$ experimental density matrix $\rho_{\rm exp}$. The leftmost subplot in Fig.~\ref{fig:qutrit_density_histogram} displays the histogram representation of $\rho_{\rm exp}$ for the two central qutrits Q$_4$Q$_5$ in the case of the initial state $\Phi_1$ with $u=0$ at $t=9$ ns. The initial reconstruction reveals significant deviations from theoretical predictions calculated using experimental parameters, with quantum state fidelity $F(\rho_{\rm exp}, \rho_{\rm th}) \sim 81.3\%$. After applying single-particle dynamical phase compensation via $Z(\theta_1)^{\dagger} Z(\theta_2)^{\dagger} \rho_{\rm exp} Z(\theta_1) Z(\theta_2)$, the calibrated density matrix $\rho'_{\rm exp}$ exhibits substantially improved fidelity of $96.5\%$ with the theoretical result. Theoretical analysis attributes this discrepancy to single-qutrit dynamical phase accumulation induced by longitudinal magnetic fields and square pulses in the measurement system. Crucially, as phase errors correspond to local unitary transformations, they do not affect key quantum correlation measures such as negativity and quantum discord.

\vspace{8mm}
\section{COMPARISON OF DYNAMICS WITH UNIFORM AND NONUNIFORM ON-SITE INTERACTIONS}
\vspace{5mm}

\begin{figure*}[b]
	\centering
	\includegraphics[width=0.99\linewidth]{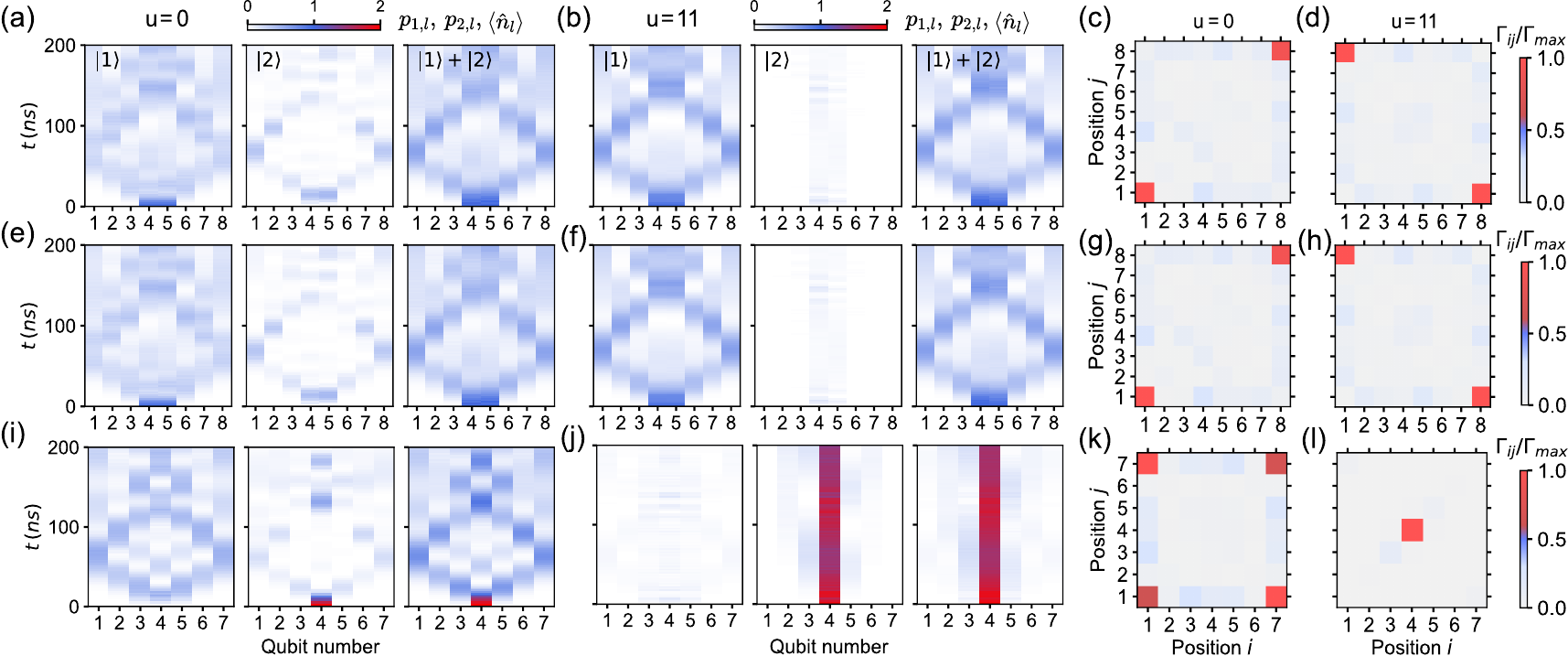}
	\caption{\textbf{Quantum walks and density-density correlations with different initial states and $u$.} Numerical simulations with experimental nonuniform $u$, calculated under the same conditions as in Fig.~2 in the main text.}
	\label{fig:theory0}
\end{figure*}

\begin{figure*}[t]
	\centering
	\includegraphics[width=0.99\linewidth]{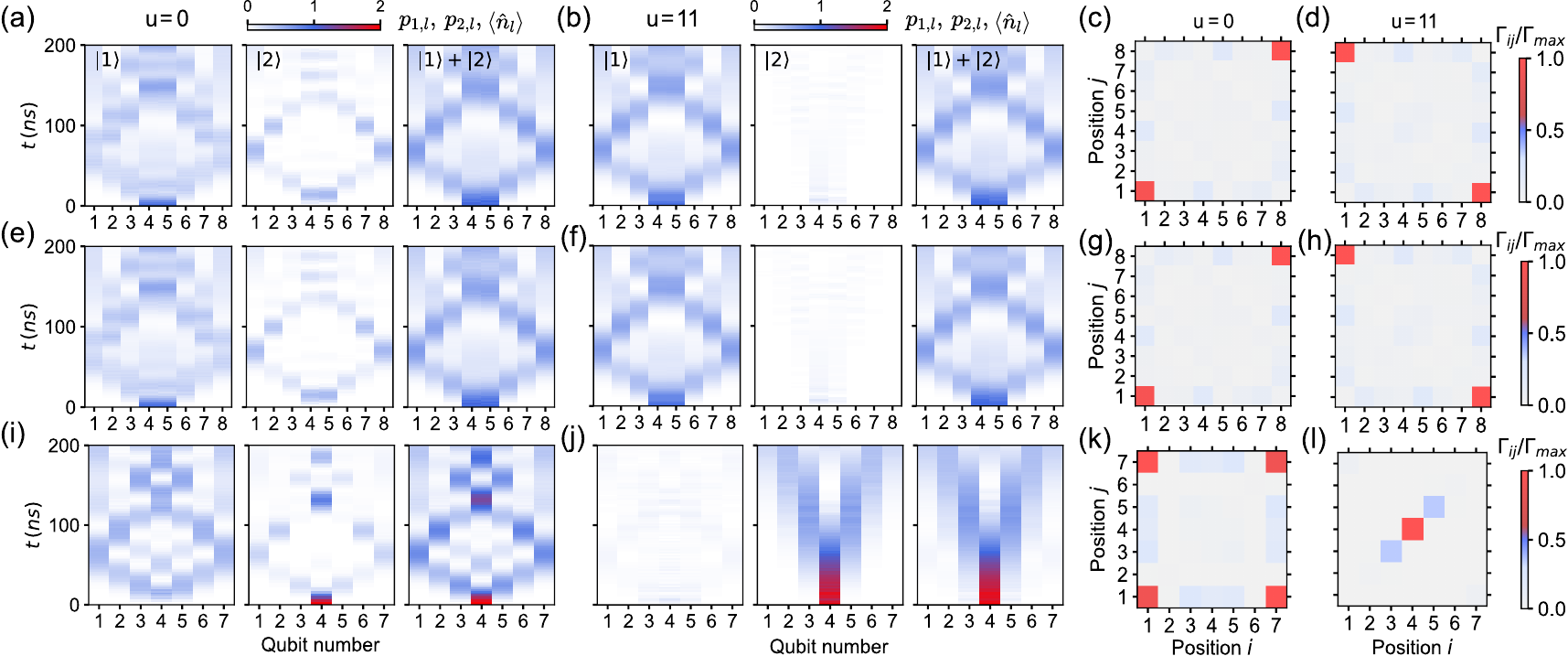}
	\caption{\textbf{Quantum walks and density-density correlations with different initial states and $u$.} Numerical simulations calculated under the same conditions as in Fig.~2 in the main text, but with ideal uniform $u$.}
	\label{fig:theory1}
\end{figure*}
\begin{figure*}[!tp]
	\centering
	\includegraphics[width=0.95\linewidth]{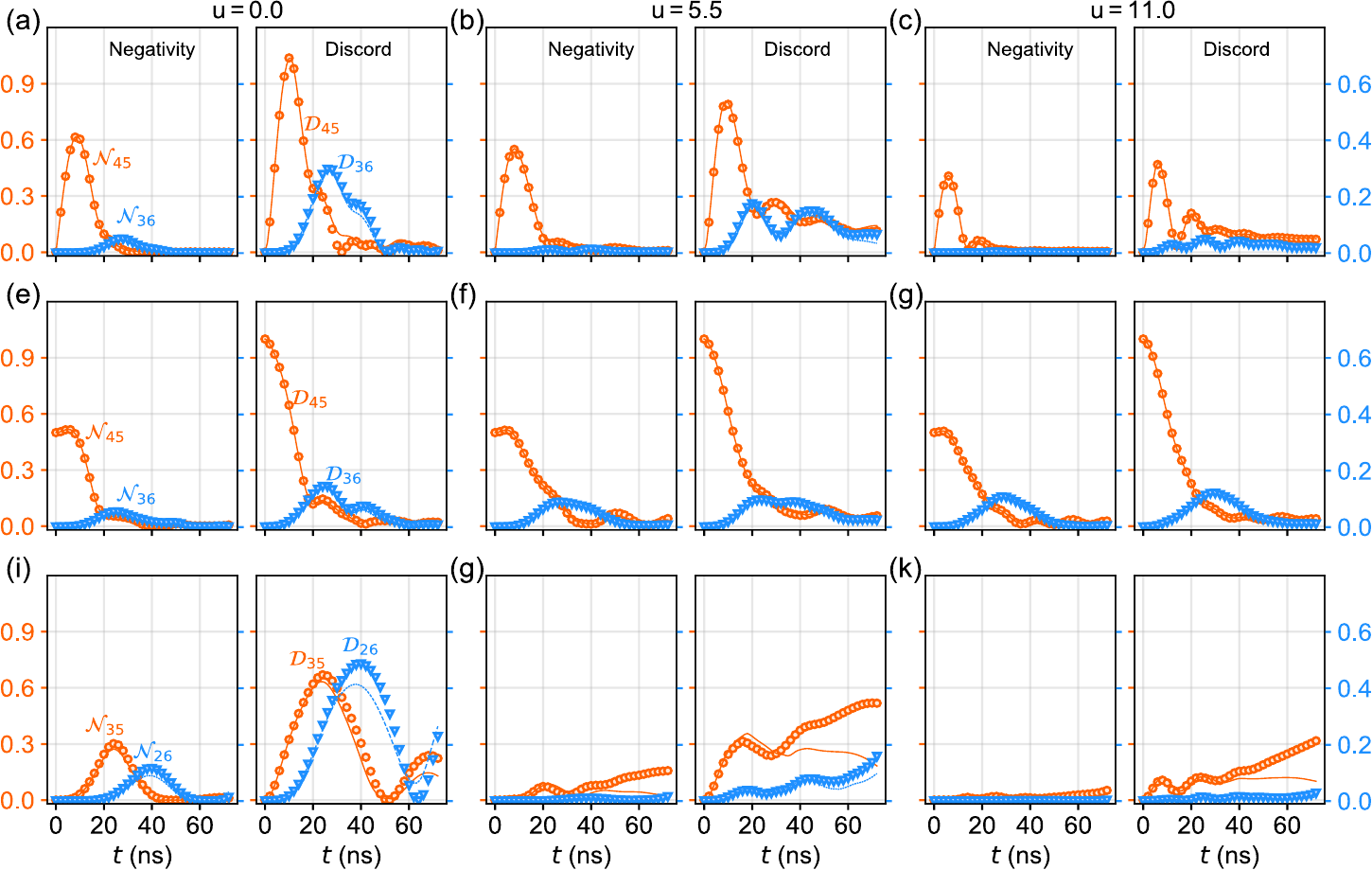}
	\caption{\textbf{Negativity and quantum discord with different initial states and $u$.} The lines are the replotted results of Fig.~3 in the main text with the same panel numbering, calculated with experimental nonuniform $u$. The symbols represent the results calculated with ideal uniform $u$ for comparison.}
	\label{fig:compare_theory}
\end{figure*}

To assess the impact of spatial nonuniformity of the engineered on-site interaction $U$, we perform numerical simulations comparing the dynamics for experimental nonuniform and ideal uniform $U$. The simulations are performed with the same initial states and interaction strengths used in the experiment.

For the experimental nonuniform $U$, we plot the corresponding simulation results in Fig.~\ref{fig:theory0}. Figures~\ref{fig:theory0}(a, b), (e, f), and (i, j) show the simulated time evolution of the particle density distribution $p_{n,l}$. The density-density correlator $\Gamma_{ij}(t) = \langle \hat{a}_i^{\dagger} \hat{a}_j^{\dagger} \hat{a}_j \hat{a}_i \rangle_t$ is presented in Figs.~\ref{fig:theory0}(c, d), evaluated at $t = 68\,\text{ns}$, when two particles are initially placed on the central two qutrits and then propagate to the chain edges. Figures~\ref{fig:theory0}(g, h) show the similar results with lower particle density when the Bell state $\ket{\Phi_2}$ is prepared at the central two qutrits. Also, Figs.~\ref{fig:theory0}(k, l) display $\Gamma_{ij}(t)$ at $t = 64\,\text{ns}$ for the case in which two particles start from a single central qutrit and subsequently spread toward the chain ends. As can be seen, these results agree well with the experimental observations in Fig.~2 of the main text. The corresponding simulation results for the ideal uniform $U$ are presented in Fig.~\ref{fig:theory1}.

These results enable a quantitative evaluation of the effect of spatial nonuniformity. As seen in Fig.~\ref{fig:theory1}, for the initial states $\ket{\Phi_1}$ and $\ket{\Phi_2}$, the dynamics of both the particle density $p_{n,l}$ and the correlator $\Gamma_{ij}(t)$ with uniform $U$ are nearly indistinguishable from the results in Fig.~\ref{fig:theory0} with nonuniform $U$. The experimental nonuniformity in $U$ (with $\Delta U \approx \pm J$) thus acts only as a weak perturbation. In contrast, the dynamics for the initial state $\ket{\Phi_3}$ show clear discrepancy. The propagation of $\ket{\Phi_3}$ in Figs.~\ref{fig:theory0}(i, j) is more suppressed compared to Figs.~\ref{fig:theory1}(i, j). This is directly reflected in the correlator $\Gamma_{ij}(t)$ in Figs.~\ref{fig:theory0}(k, l)], which remains more concentrated than that in Figs.~\ref{fig:theory1}(k, l)). The physical origin of this distinct behavior lies in the second-order tunneling process ($\sim J^2/U$) that governs the bound pair’s motion. This effective hopping rate is highly sensitive to local detuning, making it susceptible to spatial disorder in $U$, thereby suppressing particle propagation.

Figure~\ref{fig:compare_theory} compares the dynamics of negativity and quantum discord for nonuniform and uniform $U$. Consistent with the discussion above, for the initial states $\ket{\Phi_1}$ and $\ket{\Phi_2}$, the evolution of both quantities is nearly identical in the two cases, confirming the insensitivity to the spatial nonuniformity of $U$. For the initial state $\ket{\Phi_3}$, however, the results are noticeably affected by the uniformity of $U$. It appears that with increasing $U$, both negativity and quantum discord exhibit easier spatial spreading in the case of uniform on-site interactions.

\vspace{8mm}
\section{CALCULATIONS OF NEGATIVITY AND QUANTUM DISCORD}
\vspace{5mm}

Negativity serves as an operational entanglement witness for bipartite systems. Rooted in the partial transpose criterion, it is computed as 
\begin{equation}
    N(\rho) = \frac{||\rho^{PT}_{\rm AB}||_1-1}{2},
\end{equation}
where $\rho^{PT}_{\rm AB}$ is the partial transposition of state $\rho_{\rm AB}$ and $||\cdot||_1$ is the trace norm. By detecting negative eigenvalues of $\rho^{PT}_{\rm AB}$, it certifies non-separable entanglement. Negativity is utilized to quantify the temporal variation of entanglement between two qutrit subsystems in the present experiment.

Quantum discord\cite{Zurek_2001} captures non-classical correlations beyond entanglement through the mismatch between quantum mutual information 
$I(\rho_{AB})$ and classical mutual information $C(\rho_{AB})$:
\begin{equation}
    D(\rho_{AB}) = I(\rho_{AB})-C(\rho_{AB}),
\end{equation}
where $I(\rho_{AB})$ includes both classical and quantum correlations:
\begin{align}
    I(\rho_{AB}) &= S(\rho_A)+S(\rho_B)-S(\rho_{AB}).
\end{align}
Here, $\rho_A$($\rho_B$) is the reduced density matrix of particle $A$($B$), $S(\rho)$ is the von Neumann entropy of density matrix $\rho$, defined by $S(\rho)$ = $-{\rm tr}(\rho{\rm log}\rho)$ with logarithm taken to base two. $C(\rho_{AB})$ depends on the maximum information gained by measuring one particle of the total system defined as
\begin{align}
    C(\rho_{AB}) &= {\rm max}_{B_j^{\dagger}B_j}[S(\rho_A)-\sum_j q_j S(\rho_A^j)],
\end{align}
where $q_j = {\rm Tr}[B_j\rho_{AB}B_j^{\dagger}]$ is the probability of obtaining result $j$ when performing positive operator-valued measure (POVM) $\{B_j^{\dagger}B_j\}$ on subsystem B, and $\rho_A^j =  {\rm Tr}_B[B_j\rho_{AB}B_j^{\dagger}]/q_j$ is the state of subsystem $A$ after obtaining the outcome $j$.

Calculating quantum discord in a $d$-level system is a formidable computational challenge due to the necessity of optimizing over a parameter space of dimension $d(d-1)$. For our two-qutrit system, we use the method in Ref.~\cite{fu_experimental_2022}, where the measurement operators are parameterized by six parameters $\alpha, \beta, \gamma, \psi, \theta$, and $\phi$. Specifically, $\alpha, \beta \in [0,\pi]$ and $\gamma \in [-\pi/2,\pi/2]$ are used to construct a group of orthonormal bases of a qutrit system given by
\begin{align}
    {B}_p &= \begin{pmatrix}
    e^{-\mathrm{i}\phi_0} \cos\alpha \cos\beta \\
    -e^{-\mathrm{i}\gamma} \sin\beta \\
    e^{\mathrm{i}\phi_0} \sin\alpha \cos\beta
    \end{pmatrix}, \\
    {B}_0 &= \begin{pmatrix}
    e^{-\mathrm{i}\phi_0} \cos\alpha \sin\beta \\
    e^{-\mathrm{i}\gamma} \cos\beta \\
    e^{\mathrm{i}\phi_0} \sin\alpha \sin\beta
    \end{pmatrix}, \\
    {B}_m &= \begin{pmatrix}
    -e^{-\mathrm{i}\phi_0} \sin\alpha \\
    0 \\
    e^{\mathrm{i}\phi_0} \cos\alpha
    \end{pmatrix},
\end{align}
where $\phi_0 = \arctan\left( \tan\gamma \cdot \tan\left( \frac{\pi}{4} - \alpha \right) \right)$. Then parameters are used to define a general rotation of a qutrit system with the form
\begin{equation}
    R = \exp\left(-\mathrm{i}\psi S_z\right) \cdot \exp\left(-\mathrm{i}\theta S_y\right) \cdot \exp\left(-\mathrm{i}\phi S_z\right),
\end{equation}
in which 
\begin{align}
    S_y = -\frac{i}{2} \begin{pmatrix}
    0 & -1 & 0 \\
    1 & 0 & -\sqrt{2} \\
    0 & \sqrt{2} & 0
    \end{pmatrix}, \quad
    S_z = \frac{1}{\sqrt{2}} \begin{pmatrix}
    1 & 0 & 0 \\
    0 & 0 & 0 \\
    0 & 0 & -1
    \end{pmatrix}.
\end{align}
The quantum discord is obtained by traversing all the projective measurements with the basis defined above. For every two-qutrit density matrix measured in the experiments, we discretize the parameter space and use python to execute the calculations. With the increment of the discretization of the parameters, the calculation converges and then the discord is obtained.

\vspace{8mm}
\section{EXTENDED DATA}
\vspace{5mm}

\begin{figure}[htp]
    \centering
    \includegraphics[width=0.99\linewidth]{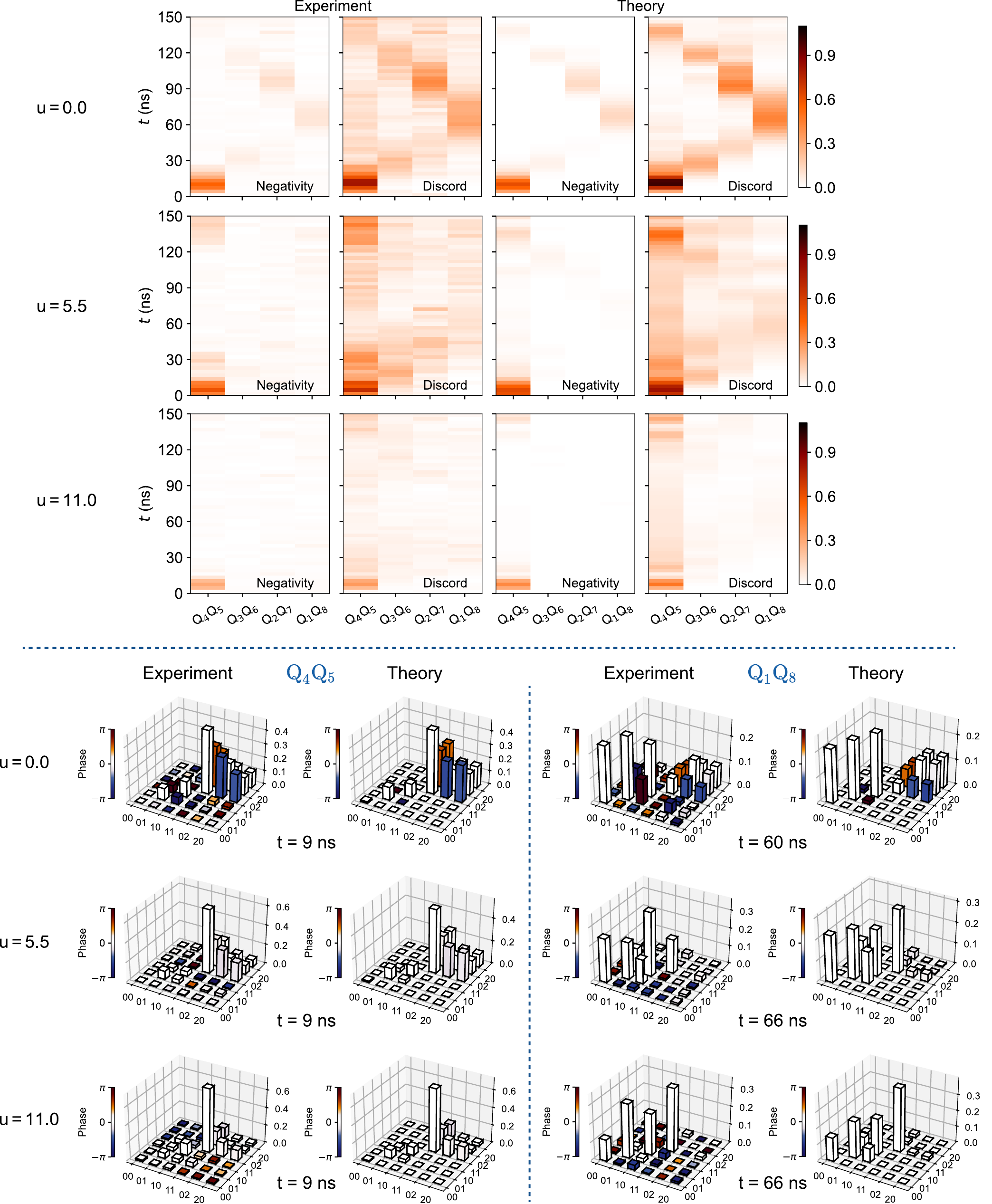}
    \caption{\textbf{Extended data for the initial state $\ket{\Phi_1}$.} Upper panels: The experimental and calculated results of negativity and quantum discord for three $u$ values indicated. Lower panels: The experimental and calculated density matrices for the central qutrit pair Q$_4$Q$_5$ (the left two columns) and boundary qutrit pair Q$_1$Q$_8$ (the right two columns) at the times indicated.}
    \label{fig:phi_1}
\end{figure}

\begin{figure}[htp]
    \centering
    \includegraphics[width=0.99\linewidth]{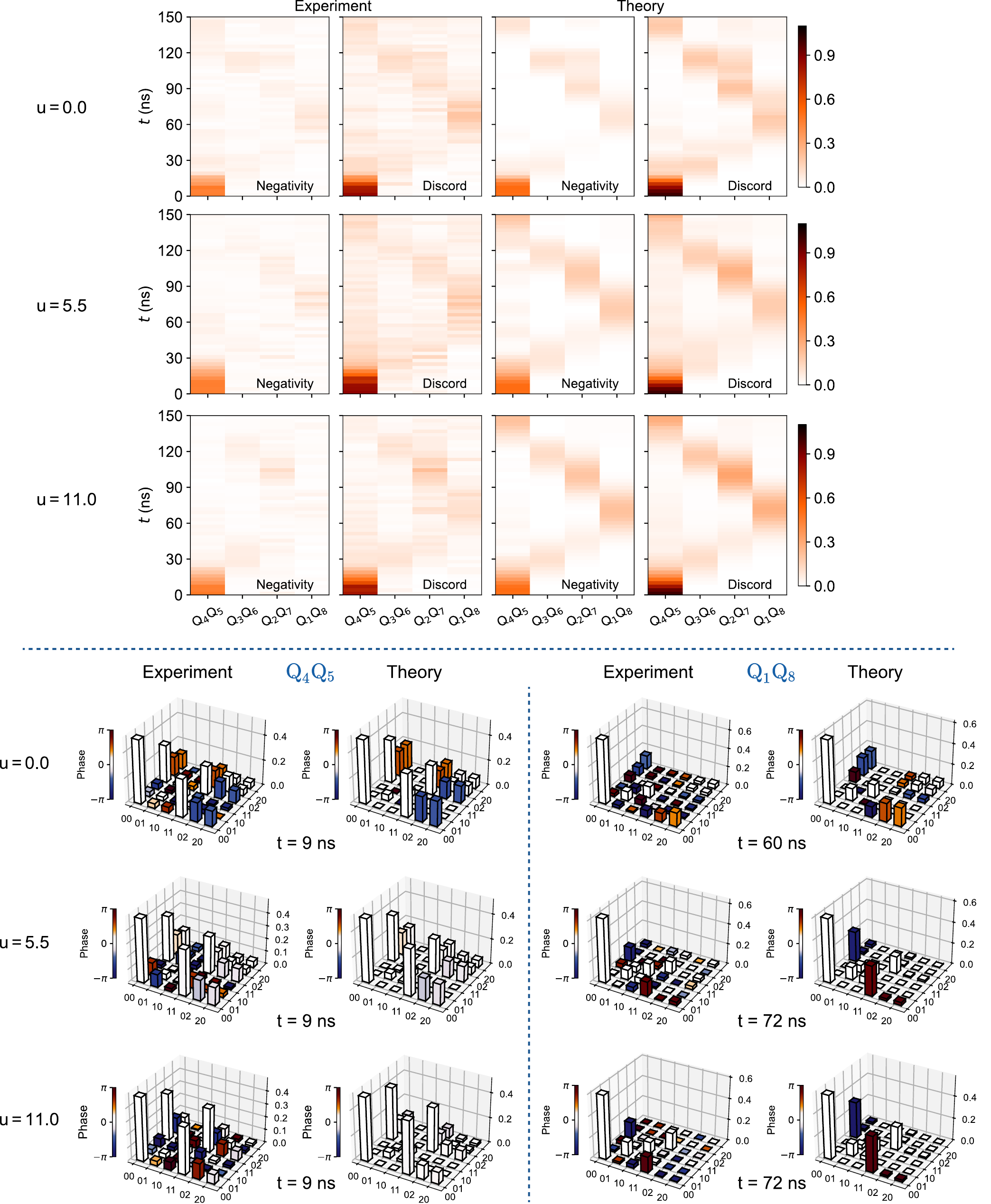}
    \caption{\textbf{Extended data for the initial state $\ket{\Phi_2}$.} Upper panels: The experimental and calculated results of negativity and quantum discord for three $u$ values indicated. Lower panels: The experimental and calculated density matrices for the central qutrit pair Q$_4$Q$_5$ (the left two columns) and boundary qutrit pair Q$_1$Q$_8$ (the right two columns) at the times indicated.}
    \label{fig:phi_2}
\end{figure}

\begin{figure}[htp]
    \centering
    \includegraphics[width=0.99\linewidth]{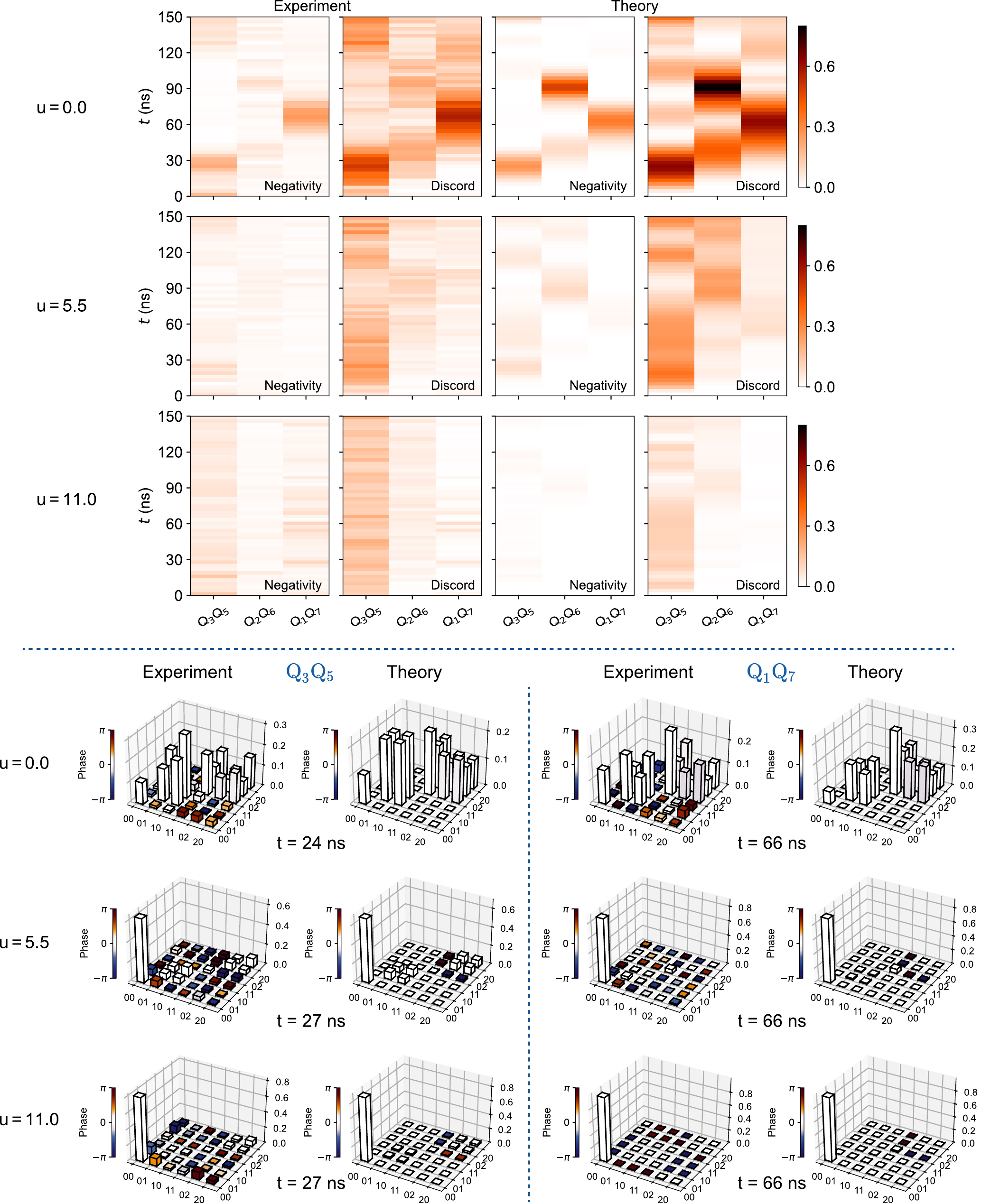}
    \caption{\textbf{Extended data for the initial state $\ket{\Phi_3}$.} Upper panels: The experimental and calculated results of negativity and quantum discord for three $u$ values indicated. Lower panels: The experimental and calculated density matrices for the central qutrit pair Q$_3$Q$_5$ (the left two columns) and boundary qutrit pair Q$_1$Q$_7$ (the right two columns) at the times indicated.}
    \label{fig:phi_3}
\end{figure}

We have measured the creation and propagation of entanglement and quantum discord for two particles walking along the entire qutrit chains and their reflections at the boundaries for the three initial states $\Phi_1$, $\Phi_2$, and $\Phi_3$ and three $u$ values of 0.0, 5.5, and 11.0. The experimental results are presented in the upper panels of Figs.~\ref{fig:phi_1}, \ref{fig:phi_2}, and \ref{fig:phi_3}, together with the theoretical results calculated using the experimental parameters. As discussed in the main text, for the three different initial states, three different forms of entanglement, namely entangled in the two-particle subspace, similar entanglement plus coherence with the $\ket{00}$ state, and entangled in both single- and two-particle subspaces, are observed. Each form is found to persist throughout the particle walking process in the case of $u \simeq$ 0. 

Let us look at the results of Fig.~\ref{fig:phi_1} for the initial state $\ket{\Phi_1}$. The lower panels show the density matrices for the central qutrit pair Q$_4$Q$_5$ and boundary qutrit pair Q$_1$Q$_8$. For $u$ = 0, we can see that the dominant form of entanglement for the central qutrit pair Q$_4$Q$_5$ at $t$ = 6 ns is similar to $\ket{\Psi_2}$ = $\left(\ket{20}+\ket{02}\pm\ket{11}\right)/\sqrt{3}$, and will later evolve to the form similar to $\ket{\Psi^{\prime}_2}$ = $\left(\ket{20}+\ket{02}\right)/\sqrt{2}$. These forms of entanglement can be clearly seen in the boundary qutrit pair Q$_1$Q$_8$ at later times, for instance, at $t$ = 57 ns when a form of entanglement close to $\ket{\Psi^{\prime}_2}$ = $\left(\ket{20}+\ket{02}\right)/\sqrt{2}$ is illustrated. These features are characteristic of quantum walks of entangled pairs. With increasing $u$, the generation of entanglement in the two-particle subspace will be increasingly suppressed due to the suppression of population transfer from single to double excitation states, together with the initial population residing in the single excitation state. In the meantime, the entanglement in the single particle subspace also grows up.

Similar conclusions can be drawn for the initial states $\ket{\Phi_2}$, as can seen in Figs.~\ref{fig:phi_2}. In this case, clear signature of the Bell-state entanglement can be seen throughout particle walks in the chain for all $u$ values, since the double excitation state is not involved in the Bell state.

The results for the initial states $\ket{\Phi_3}$ are shown in Figs.~\ref{fig:phi_3}. The form of entanglement in both the single- and two-particle subspaces can be seen, which propagates and persists during particle walks for $u \simeq$ 0. With increasing $u$, both the generation and propagation of entanglement become increasingly small in the time scale shown. Over a much larger time scale, the entanglement in the two-particle subspace is expected, but that in the single-particle subspace will be significantly suppressed due to the suppression of population transfer from double to single excitation states, together with the initial population residing in the double excitation state.

\newpage


%